\documentclass[journal,twocolumn,10pt]{IEEEtran}
\usepackage{graphicx}
\usepackage{verbatim}
\usepackage{upgreek}
\usepackage{amssymb,amsmath}
\usepackage{color}
\usepackage{epstopdf}
\usepackage{bm}
\usepackage{cite}
\usepackage{array,color}
\usepackage{algorithm}
\usepackage{algorithmicx}
\usepackage{algpseudocode}
\usepackage{amsmath}
\usepackage{amsfonts}
\usepackage{amsthm}
\usepackage{graphics}
\usepackage{epsfig}
\usepackage{soul}
\usepackage{booktabs}
\usepackage{multirow}
\usepackage{stfloats}
\soulregister\cite7
\soulregister\ref7
\usepackage{subfigure}
\usepackage{tabularx}
\usepackage{makecell}
\usepackage{graphicx}
\usepackage{gensymb}
\renewcommand\appendix{\setcounter{secnumdepth}{3}}
\newtheorem{lemma}{Lemma}

\usepackage{etoolbox}
\newcommand{\tabincell}[2]{\begin{tabular}{@{}#1@{}}#2\end{tabular}} 
\makeatletter
\patchcmd{\@makecaption}
{\scshape}
{}
{}
{}
\makeatletter
\patchcmd{\@makecaption}
{\\}
{.\ }
{}
{}
\makeatother

\ifCLASSINFOpdf
\else
\fi

\begin{document}

\title{\Huge Joint Offloading and Beamforming Design in Integrating Sensing, Communication, and Computing Systems: A Distributed Approach}

\author{Peng Liu, Zesong Fei,~\IEEEmembership{Senior~Member,~IEEE}, Xinyi Wang,~\IEEEmembership{Member,~IEEE}, Jingxuan Huang,  \\ Jie Hu, ~\IEEEmembership{Member,~IEEE}, and J. Andrew Zhang,~\IEEEmembership{Senior~Member,~IEEE}
	
	\thanks{Peng Liu, Zesong Fei, Xinyi Wang, and Jingxuan Huang are with the School of Information and Electronics, Beijing Institute of Technology, Beijing 100081, China (E-mail:bit\_peng\_liu@163.com, feizesong@bit.edu.cn, bit\_wangxy@163.com, jxhbit@gmail.com).}
\thanks{Jie Hu is with the School of Information and Communication Engineering, University of Electronic Science and Technology of China, Chengdu 611731, China (E-mail: hujie@uestc.edu.cn).}
	\thanks{J. Andrew Zhang is with the School of Electrical and Data Engineering, University of Technology Sydney, NSW, Australia 2007. (E-mail:Andrew.Zhang@uts.edu.au)}
	
}

\maketitle

\begin{abstract}
\textcolor{black}{When applying} integrated sensing and communications (ISAC) \textcolor{black}{in future mobile networks, many sensing tasks have low latency requirements, preferably being implemented at terminals. However, terminals often have limited computing capabilities and energy supply.} In this paper, we investigate the effectiveness of leveraging the advanced computing capabilities of mobile edge computing (MEC) servers and the cloud server to address the sensing tasks of ISAC terminals. Specifically, we propose a novel three-tier integrated sensing, communication, and computing (ISCC) framework composed of one cloud server, multiple MEC servers, and multiple terminals, where the terminals can optionally offload sensing data to the MEC server or the cloud server. \textcolor{black}{The offload message is sent via the} ISAC waveform, \textcolor{black}{whose echo is used for sensing.} We jointly optimize the computation offloading and beamforming strategies to minimize the average execution latency while satisfying sensing requirements. In particular, we propose a low-complexity distributed algorithm to solve the problem. Firstly, we use the alternating direction method of multipliers (ADMM) and derive the closed-form solution for offloading decision variables. Subsequently, we convert the beamforming optimization sub-problem into a weighted minimum mean-square error (WMMSE) problem and propose a fractional programming based algorithm. Numerical results demonstrate that the proposed ISCC framework and distributed algorithm significantly reduce the execution latency \textcolor{black}{and the energy consumption} of sensing tasks at a lower computational complexity compared to existing schemes.

\vspace{2ex}
\textbf{Keywords: Integrated sensing and communication, mobile edge computing, alternating direction method of multipliers, computation offloading, beamforming design.} 
\end{abstract}

\IEEEpeerreviewmaketitle
\section{Introduction}
The forthcoming sixth-generation (6G) wireless networks are anticipated to serve as a fundamental catalyst for numerous emerging applications, including unmanned aerial vehicles (UAV), connected autonomous systems, extended reality, and digital twins \cite{2020NW}. These applications require wireless networks to deliver high-accuracy sensing capabilities and high-quality wireless connectivity. As communication and sensing systems advance toward larger antenna arrays and higher frequency bands, their increasing similarities in hardware and signal processing perspectives have presented a clear opportunity for integrated sensing and communication (ISAC) \cite{CS2021}.

With the evolution of the multiple-input and multiple-output (MIMO) technique, beamforming method in ISAC systems has garnered widespread research interest, which simultaneously achieves sensing and communication functions through different spatial beams \cite{Liu2018TWC,xv2024,TSP2020,2024cui,wclliu2022,ren2023,wang2021}. The author in \cite{Liu2018TWC} explored the deployments of two antenna arrays within ISAC systems and showed the superior performance of the shared deployment compared to the separate deployment. Furthermore, the authors in \cite{xv2024} delved into optimal spatial waveform design in ISAC system. In addition, a novel ISAC beamforming approach was introduced in \cite{TSP2020}, demonstrating that joint beamforming design can achieve similar estimation performance compared to sensing-only design. For moving sensing and communication targets, \textcolor{black}{the authors in \cite{2024cui} proposed an extended Kalman filter based ISAC beamforming method for angle prediction and assist beam alignment.} Furthermore, \textcolor{black}{to protect the information from being eavesdropped, the authors in \cite{wclliu2022} and \cite{ren2023} investigated ISAC beamforming design while embedding artificial noise into ISAC waveform.} To address the communication performance deterioration due to the decreased degrees of freedom (DoFs) constrained by sensing requirements, the authors in \cite{wang2021} studied joint reflection and waveform design in an intelligent reconfigurable surface (IRS)-aided ISAC system, effectively mitigating multi-user interference by reconfiguring the propagation environments with IRS.

Although the ISAC technique offers numerous applications and advantages, it also faces challenges in computational complexity and latency. In particular, the sensing process generates a significant number of echo signals, while some advanced sensing signal processing methods, such as compressed sensing (CS) \cite{radar3}, spatial spectrum estimation (SSE) \cite{radar5}, continuous wavelet transform (CWT) \cite{radar4}, \textcolor{black}{and synthetic aperture radar for aerial platforms \cite{sar2012}}, are typically intricate, requiring substantial computational resources. Although some artificial intelligence methods, such as reinforcement learning and deep learning \cite{computing1,computing2}, have shown their advanced capabilities of high-resolution sensing, they are typically complicated and may also impose severe computational burden for resource-limited ISAC terminals. Fortunately, mobile edge computing (MEC) has been identified as a crucial technique for alleviating terminal computational burdens by offloading computationally intensive and latency-sensitive tasks to the network edge \cite{MEC,CSMEC1,CSMEC2}. There have been numerous research efforts focusing on employing the MEC technique to address the computational issue in ISAC systems. Liu \textit{et al.} \cite{TGCN2023} proposed a joint computation offloading and resource allocation strategy for integrating sensing, communication, and computing (ISCC)-aided vehicle-to-everything (V2X) networks, which improves the energy efficiency for completing the sensing data fusion tasks  while ensuring the execution latency requirement. The authors in \cite{UAVMEC} investigated the issues of trajectory optimization and power allocation in the ISCC-aided UAV system with the aim of minimizing the UAV energy consumption and sensing data collection time while satisfying radar sensing accuracy. 

{  
	\begin{table*}[!ht]
		\renewcommand{\arraystretch}{1.3}
		\caption{Comparison between our work and the existing ISCC works}
		\label{1}
		\centering
		\renewcommand{\arraystretch}{1.3}
		\begin{tabular}{|c|c|c|c|c|c|c|c|c|}
			\hline
			\bfseries Reference  & \tabincell{c}{\bfseries ISAC} &  \tabincell{c}{\bfseries Mobile edge \\ \bfseries computing} & \tabincell{c}{\bfseries Mobile cloud \\ \bfseries computing} & \tabincell{c}{\bfseries Multi-MEC servers} & \tabincell{c}{\bfseries MIMO } & \tabincell{c}{\bfseries Optimize latency }& \tabincell{c}{\bfseries Distributed }\\
			\hline
			\cite{TGCN2023} & \checkmark &\checkmark & & \checkmark& &\checkmark & \\
			\hline
			\cite{UAVMEC} &\checkmark&\checkmark && &&\checkmark& \\
			\hline
			\cite{wcl2023} &\checkmark&\checkmark & &\checkmark&\checkmark & & \\
			\hline
			\cite{TWC2023} &\checkmark&\checkmark& & &\checkmark&\checkmark& \\
			\hline
			\cite{ISCC2} &\checkmark&\checkmark& & &\checkmark & &  \\
			\hline
			\cite{Huang2024} &\checkmark&\checkmark& & &\checkmark & &  \\
			\hline
			\cite{ISCC1} &\checkmark&\checkmark& & &\checkmark & &  \\
			\hline
			Our work &\checkmark&\checkmark &\checkmark&\checkmark&\checkmark&\checkmark&\checkmark \\
			\hline
		\end{tabular}	
\end{table*}}

The above works primarily focused on the single-antenna system. To utilize spatial DoFs brought by the MIMO technique, the authors in \cite{wcl2023} studied the beamforming optimization in ISCC systems to minimize the energy consumption of terminals under offloading throughput and processing delay constraints. Furthermore, the authors in \cite{TWC2023} formulated the inter-beam interference model for the ISCC systems and optimized the offloading strategy as well as power allocation schemes to reduce task execution latency. The authors in \cite{ISCC2} studied the beamforming optimization and computational resource allocation in MIMO-aided ISCC systems, constructing a multi-objective problem concerning computational offloading energy consumption and sensing pattern design. In \cite{Huang2024}, the authors proposed a paradigm of MEC-aided ISAC with short-packet transmissions and optimized the size and duration of each short packet to minimize energy consumption. To further enhance the computational efficiency of the ISCC system, the authors in \cite{ISCC1} introduced the over-the-air federate learning (AirFL) \textcolor{black}{framework, where the terminals train local computing models based on raw data and upload them to a central server via over-the-air computation for model aggregation.} However, most prior works have only considered a two-tier ISCC framework comprising one MEC server, and have not fully leveraged multiple potential MECs and cloud servers with more computational resources.  
Compared to the two-layer architecture with a single MEC server \cite{TGCN2023,UAVMEC,wcl2023,TWC2023,ISCC2, ISCC1}, the cloud-edge-terminal three-tier architecture can offer more task execution DoFs to reduce task execution latency, yet it also introduces more complexity for the design of offloading strategies and ISAC beamforming.

Motivated by the aforementioned analysis on the limits of existing schemes, in this paper, we explore a novel multi-user MIMO ISCC framework comprising one mobile cloud computation (MCC)  server, multiple MEC servers, and numerous ISAC terminals. There are three strategies for processing the sensing tasks of terminals, including local execution, MEC server execution, and MCC execution. We aim to minimize the average sensing task execution latency by jointly optimizing computation offloading and beamforming strategies. The differences between our work and existing ISCC works are summarized in Table {\ref{1}}, and the main contributions are summarized as follows:
\begin{itemize}
\item{We introduce a three-tier ISCC framework composed of one MCC server, multiple MEC servers, and multiple terminals. Utilizing the ISAC beamforming technique, each terminal can optionally offload sensing data to either MEC server or MCC server for processing while \textcolor{black}{using the echoes of the offload signals for sensing.} Furthermore, we formulate an optimization problem to minimize the average task execution latency at terminals by jointly designing offloading decisions and beamforming variables.}
\item\textcolor{black}{To address the discrete and non-convex coupled optimization problem with a lower complexity,} we propose a distributed optimization framework for joint offloading and beamforming design. Specifically, we utilize the block coordinate descent (BCD) technique to decompose the non-convex problem with coupled variables into two sub-problems and iteratively solve them. 
\item \textcolor{black}{For the discrete computation offloading optimization sub-problem, we integrate the Alternating Direction Method of Multipliers (ADMM) with continuous relaxation and expansion techniques to develop a novel distributed algorithm \textcolor{black}{that effectively addresses} integer programming challenges. To reduce the complexity of the algorithm, we derive closed-form solutions for the offloading decision variables using the Karush-Kuhn-Tucker (KKT) conditions. For the non-convex ISAC beamforming optimization sub-problem, we develop a distributed algorithm that combines fractional programming (FP), weighted minimum mean square error (WMMSE), and successive convex approximation (SCA) techniques. We first employ the FP technique to transform the original subproblem into a more manageable form. Next, we decompose it based on the WMMSE transformation for parallel processing. Finally, we apply the SCA technique to effectively tackle the non-convex sensing constraints.}
\item{Numerical results are provided to demonstrate the effectiveness of the proposed three-tier ISCC scheme in executing sensing tasks. Compared to existing schemes, the proposed distributed algorithm is shown to be able to reduce task  execution latency and energy consumption significantly with lower complexity. \textcolor{black}{Additionally, we illustrate the trade-offs within the three-tier ISCC system between target sensing and data offloading, and those between the  computational resources at MEC servers and sensing performance. \textcolor{black}{By strategically prioritizing execution speed, sensing tasks can be completed with reduced latency and more efficient use of computational resources, while still maintaining a balanced level of sensing performance.} }
}
\end{itemize}

The remaining sections of this paper are organized as follows. In section II, we introduce the system model, encompassing both the signal and computation models, and outlines the formulation of the optimization problem. Section III proposes a distributed solution approach for the joint optimization problem concerning  offloading and ISAC beamforming strategies. In section IV, the simulation results are presented. Finally, Section V concludes the paper with a summary of our findings and contributions.

\begin{table}[!t]
	\renewcommand{\arraystretch}{1.3}
	\caption{Summary of Symbols}
	\label{table_example}
	\centering
	\begin{tabular}{l l}
		\hline
		\bfseries Symbols & \multicolumn{1}{c}{\bfseries Definition}\\ 
		\hline
		$\mathbf{x}_k$ & \tabincell{l}{Terminal $k$'s transimit signal for sensing and offloading}\\
		$\mathbf{w}_k$ & \tabincell{l}{Terminal $k$'s ISAC beamforming vector}\\
		$s_k$ &  \tabincell{l}{Terminal $k$'s communication stream}\\
		$\mathbf{H}_{lk}$ & \tabincell{l}{Channel matrix from terminal $k$ to MEC server $l$}\\
		$\mathbf{H}^I_{kj}$ & The interference channel from terminal $j$ to terminal $k$\\
		$d_{l,k}$ & \tabincell{l}{The distance from terminal $k$ to MEC server $l$}\\
		$d^t_{k}$ & \tabincell{l}{The distance from terminal $k$ to the sensing target}\\
		
		$\mathbf{n}_{l}$ & \tabincell{l}{The comlex Gaussian noise at MEC server $l$}\\
		$\mathbf{a}({\theta}_k)$ & \tabincell{l}{Terminals' array steering vector}\\
		$B$ &  \tabincell{l}{ISCC signal bandwidth} \\
		${\gamma}^r_{k}$ & \tabincell{l}{The radar echo SINR of terminal $k$'s target}\\
		$\rho_0$ &  The path loss at the reference distance $d_0 = 1$ m \\
		$\xi_k$ &  The radar RCS of terminal $k$'s target \\
		$Z_k$ & The number of bits in sensing task ${C}_k$\\ 
		$S_k$ &  The CPU cycles required to accomplish computation of ${C}_k$\\
		$\beta$ &  \tabincell{l}{ The task computation indensity}\\
		$f_{k}^L$ & \tabincell{l}{Terminal $k$'s CPU frequency}\\
		$f_{k}^E$ & \tabincell{l}{The CPU frequency allocated to terminal $k$ by the MEC}\\
		$f_{k}^C$ & \tabincell{l}{The CPU frequency allocated to terminal $k$ by the MCC}\\
		$T_{k}^L$ & \tabincell{l}{The execution latency for locally processing the sensing task}\\
		$T_k^E$ &\tabincell{l}{The execution latency for offloading task to the MEC server}\\
		$T_k^C$  & \tabincell{l}{The execution latency for offloading task to the MCC server}\\
		$p_{k}^L$ & \tabincell{l}{The power consumption for locally processing the sensing task}\\
		$p_{k}^E$ & \tabincell{l}{The power consumption for offloading task to the MEC server}\\
		$p_{k}^C$ & \tabincell{l}{The power consumption for offloading task to the MCC server}\\
		${G}_l$ & The maximum computation capability of MEC server $l$\\
		$P_{th}$ & \tabincell{l}{The terminal's maximum power}\\
		$R_{lk}$ & \tabincell{l}{The uplink transmit rate of from terminal $k$ to MEC server $l$}\\
		$\eta$  & The power coefficient related to the chip architecture\\
		$r^f$  & The transmission rate between MEC server and MCC server\\
		$\Gamma_{r}$ & The sensing SINR threshold of terminals \\
		\hline
	\end{tabular}
\end{table}

\textit{Notations:} The uppercase boldface and lowercase boldface denote matrices and vectors, respectively;  det($\cdot$) and $\text{tr}(\cdot)$ denote the determinant and the trace of the matrix, respectively; $[\cdot]^H$, $[\cdot]^*$ $[\cdot]^{-1}$ denote the conjugate-transpose and conjugate operations, respectively; $||\cdot||$ is the L2 norm of vectors; $\mathbf{I}_M$ is an  $M\times M$ identity matrix; $\mathbb{E}[\cdot]$ and $\mathbb{C}$ denote the expectation operation and set of complex numbers, respectively. The other notations used in the paper are summarized in Table II.

\begin{figure}[!t]
	\centering
	\includegraphics[width=3.5in]{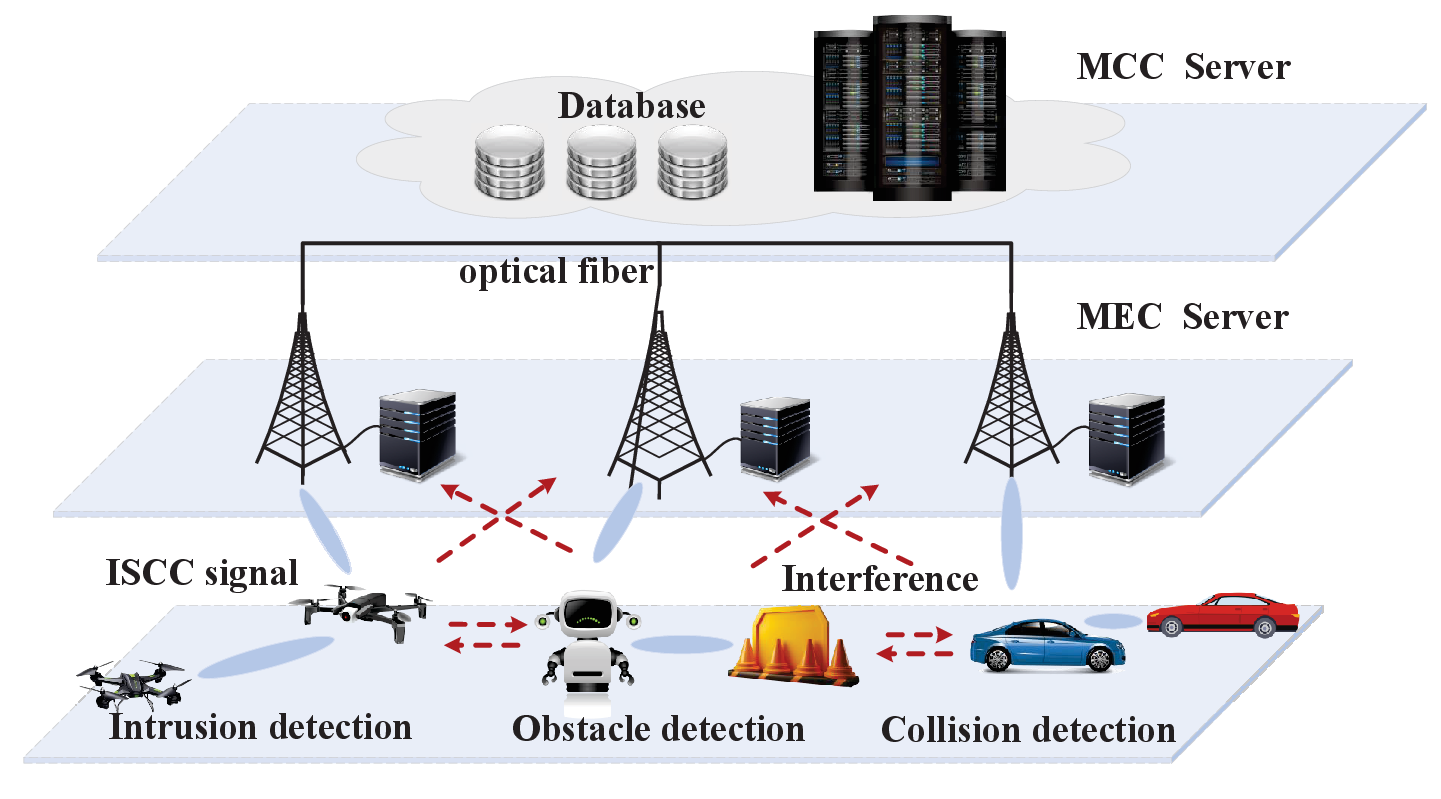}
	\caption{An illustration of the three-tier network architecture comprising a CET ISCC framework.}
	\label{fig:f0}
\end{figure}


\section{System Model}

As shown in Fig. {\ref{fig:f0}}, \textcolor{black}{we consider a Cloud-Edge-Terminal (CET) three-tier network architecture consisting of one MCC server, $L$ $M$-antenna base stations (BS), each equipped with one MEC server, denoted by $\mathcal{L}=\{1,\cdots,L\}$ and $K$ ISAC terminals, denoted by $\mathcal{K}=\{1,\cdots,K\}$.} When performing sensing tasks, such as collision detection and target identification, a large amount of computation-intensive sensing data will be generated, while the computing capacity of the terminals is generally limited, resulting in a long processing latency. To addresss this issue, the terminals optionally offload the heavy sensing tasks to the MEC or MCC server. To enable simultaneous ISAC signal transmission and echo signal reception, each terminal is equipped with $N_t$ transmit antennas and $N_r$ receive antennas, and space-division multiple access (SDMA) instead of orthogonal frequency division multiple access (OFDMA) is employed such that the spectral efficiency and sensing accuracy can be improved with a large bandwidth. Without loss of generality, we assume $N_t=N_r=N$. In general, the computation task offloading strategy can be categorized into two types, i.e.,  partial and binary computation offloading \cite{MEC}. The former assumes the input bits of computation tasks are bitwise independent such that they can be split into different parts and computed in parallel. On the other hand, the binary computation offloading is more suitable for tasks that are highly integrated and cannot be partitioned. In this paper, due to the holistic nature of the sensing signal processing task, we consider the binary offloading mode.

\textbf{Remark 1.} \textit{\textcolor{black}{In the considered ISCC model, there are two main reasons for utilizing SDMA instead of OFDMA. Firstly, OFDMA typically allocates non-contiguous spectral resources to different terminals, which can severely impair their sensing performance due to the limited bandwidth. In contrast, SDMA leverages the spatial-domain dimension, allowing terminals to utilize the entire bandwidth and achieve higher range resolution. Secondly, for mobile nodes, OFDMA is highly sensitive to Doppler shifts and suffers from severe inter-user interference due to disruptions in orthogonality in dynamic scenarios. SDMA, however, leverages spatial separation \textcolor{black}{properties} and is less impacted by Doppler shifts. Additionally, it can dynamically adjust its beamforming vectors to adapt to changes in node positions, making it more suitable for mobile environments.}}

\subsection{ISCC Signal Model}
The ISCC signal transmitted by the $k$-th terminal is expressed as
\begin{equation}
\mathbf{x}_{k} = \mathbf{w}_{k}s_{k},
\end{equation}
where $s_{k}$ denotes terminal $k$'s communication symbol to be uploaded to the MEC server, and $\mathbf{w}_{k}\in\mathbb{C}^{N\times 1}$ denotes the ISAC beamforming vector. We assume that each terminal can only select one MEC/MCC server for  computation offloading and all terminals operate within the same frequency band, thereby resulting in multi-terminal interference (MTI) at MEC servers. The signal received by MEC server $l$ is expressed as
\begin{align}
\label{yl} 
\mathbf{y}_{l}&=\mathbf{H}_{lk}^{H}\mathbf{x}_{k}+\sum_{i=1,i\neq k}^K\mathbf{H}_{li}^{H}\mathbf{x}_{i} +\mathbf{n}_l ,
\end{align}
where $\mathbf{H}_{lk}=\sqrt{\rho_0 d_{l,k}^{-2}} \tilde{\mathbf{H}}_{lk}\in\mathbb{C}^{N \times M}$ denotes the channel matrix from terminal $k$ to MEC server $l$, $\rho_0$ represents the path loss at the reference distance $d_0 = 1$ m, $d_{l,k}$ is the distance from terminal $k$ to MEC server $l$, and $\tilde{\mathbf{H}}_{lk}$ denotes the small-scale fading components following complex Gaussian distribution with zero mean and unit variance. 
$\mathbf{n}_{l}\sim \mathcal{CN}(0, {\sigma_b}^2\mathbf{I}_M)$ represents the additive white Gaussian noise (AWGN) at MEC server $l$. \textcolor{black}{We assume all the channels are block-faded, quasi-static, and thus can be perfectly estimated through orthogonal pilot sequences \cite{Liu2018TWC}. Thereafter, the echoes of large-volume data-bearing signals are used for sensing via the ISAC technique.}

 According to (\ref{yl}), The uplink transmission rate from terminal $k$ to MEC server $l$ is expressed as
\begin{equation}
\label{r1}
R_{lk} = B\log\det(\mathbf{I}_M+\mathbf{H}_{lk}^{H}\mathbf{w}_{k}\mathbf{w}_{k}^H\mathbf{H}_{lk}\mathbf{D}^{-1}_{lk} ),
\end{equation}
where $B$ is the signal bandwidth of terminals. $\mathbf{D}_{lk}$ is the covariance matrix of MTI and noise, which is given as
\begin{align}
\mathbf{D}_{lk} ={\sum_{i=1,i\neq k}^K\mathbf{H}_{li}^{H}\mathbf{w}_{i}\mathbf{w}^H_{i}\mathbf{H}_{li}}+{\sigma_b}^2\mathbf{I}_M,
\end{align}

On the other hand, the echo signal received at terminal $k$ is expressed as
\begin{align}
\mathbf{y}_k^{r}&=\zeta_k\mathbf{a}^*({\theta}_k) \mathbf{a}^H({\theta}_k) \mathbf{x}_k+\sum_{j=1, j \neq k}^K \mathbf{H}^{I}_{kj} \mathbf{x}_j+\mathbf{n}_k,
\end{align}
where $\mathbf{H}^{I}_{kj}=\sqrt{\rho_0 d_{k,j}^{-2}} \tilde{\mathbf{H}}^{I}_{kj}\in\mathbb{C}^{N \times N}$ denotes the interference channel from terminal $j$ to terminal $k$ with $d_{k,j}$ being the distance from terminal $j$ to terminal $k$ and $\tilde{\mathbf{H}}^{I}_{kj}$ being the small fading components, $\mathbf{n}_k\sim \mathcal{CN}(0, {\sigma_k}^2\mathbf{I}_N)$  denotes the AWGN at terminal $k$, and $\zeta_k$ contains the impacts of the path loss and complex reflection coefficients, which is expressed as \cite{liu2023}
\begin{equation}
{\zeta}_k = \sqrt{{\rho_0}/{{d^t_{k}}^{2}} \times {\xi}_k/{{d^t_{k}}^{2}}},
\end{equation}
where ${\xi}_k$ is the radar cross-section (RCS) of terminal $k$'s target, $d^t_{k}$ is the distance from terminal $k$ to the sensing target, and $\mathbf{a}({\theta}_k)$ denotes the steering vector as follows.
\begin{equation}
\mathbf{a}(\theta_k) = [1, {\rm e}^{j2\pi \alpha \sin(\theta_k)}, \cdots, {\rm e}^{j2\pi (N -1) \alpha \sin(\theta_k)}]^T \in \mathbb{C}^{N \times 1},
\end{equation}
where $\alpha$ represents the normalized interval between adjacent antennas and $\theta_k$ represents the angle of arrival (AoA) corresponding to the terminal $k$'s sensing target. In this paper, we employ the echo SINR as the sensing performance metric, which is written as
\begin{equation}
\label{prob} 
{\gamma}^r_{k} =\frac{\zeta_k^2\text{tr}(\mathbf{A}({\theta}_k)\mathbf{w}_{k}\mathbf{w}^H_{k}\mathbf{A}({\theta}_k)^H)} {\text{tr}({\sum_{j=1,j\neq k}^K{{\mathbf{H}}_{kj}^{IH}}\mathbf{w}_{j}\mathbf{w}^H_{j}{\mathbf{H}}^{I}_{kj}})+\sigma_k^2},
\end{equation}
where $\mathbf{A}({\theta}_k)=\mathbf{a}^*({\theta}_k) \mathbf{a}^H({\theta}_k)$. \textcolor{black}{It can be observed from (\ref{prob}) that for improving the sensing accuracy, it would be better to align the \textcolor{black}{beam} towards sensing targets while minimizing interference to other terminals. However, since the ISAC waveform is also utilized to offload sensing tasks to MEC servers, the beamforming vectors also affect the offloading rate, \textcolor{black}{as will be quantitatively analyzed next.} Consequently, the trade-off between sensing accuracy and execution latency must be addressed.}
\subsection{Computation and Cost Models}
We consider a widely adopted computational task model ${C}_k=({Z}_k,{S}_k)$ in MEC and MCC studies \cite{MEC}, where ${Z}_k$ (bit/s) represents the number of bits in sensing task ${C}_k$ and ${S}_k$ is the central processing unit (CPU) cycles required to accomplish computation of the sensing task ${C}_k$. Referring to \cite{ISCC2}, the size of ${Z}_k$ is determined by the engineering implementation and radar signal processing model, which is linearly correlated with sampling frequency. According to the Nyquist sampling theorem, we assume that ${Z}_k$ is also linearly related with the signal bandwidth $B$.  

To characterize the offloading strategy, i.e., local execution, MEC server execution, MCC server execution, adopted by each terminal, we define the binary offloading decision variables $\bm{a}_l=\{a_{lk},a_{l2}\cdots,a_{lK}\}, a_{lk}\in\{0,1\},  \forall k$ and $\bm{b}_l=\{b_{11},b_{l2}\cdots,b_{lK}\}, b_{lk}\in\{0,1\},  \forall k$, respectively. When $a_{lk}=1, b_{lk}=0$, the sensing task is offloaded to MEC server $l$ for computation, and  when $a_{lk}=0, b_{lk}=1$, terminal $k$'s sensing task is offloaded to the MCC server through MEC server $l$. As each terminal can only select one offloading strategy, we have
\begin{equation}
\sum_{l=1}^La_{jk}+\sum_{l=1}^Lb_{lk}\leq 1, \forall k\in \mathcal{K}.
\end{equation}

Note that different offloading strategies result in different processing latency and power consumption. \textcolor{black}{It is worth noting that the sizes of signalling overhead and results are usually much smaller than that of the original sensing echo signals; thus, the corresponding feedback latency is considered to be negligible \cite{JMEC}, \cite{JMEC2}. Specifically, the considered latency cost mainly includes:
\begin{itemize}
	\item The execution latency for local computing, on MEC server, and on MCC server;
	\item The task offloading latency from terminals to MEC server;
	\item The backhaul transmission latency from MEC server to MCC server.
\end{itemize}}

\textcolor{black}{Considering that MEC and MCC server are generally powered by the power grid, only the power consumption at terminals is of concern \cite{wcl2023}. Therefore, the considered power consumption cost mainly includes those for local computing and for ISAC signal transmission.}
	
Next, we discuss the computation overhead of the three strategies, respectively.
\subsubsection{Local Execution}
For the local execution strategy, i.e., $a_{lk}=0, b_{lk}=0$, terminal $k$ executes its sensing computation task ${C}_k$ locally. The execution latency for locally  processing the sensing task ${C}_k$ is expressed as
\begin{equation}\label{t1}
T_k^L=\frac{{S}_k}{f^L_k}=\frac{\beta{Z}_k}{f^L_k}, ~\forall k\in \mathcal{K},
\end{equation}
where $f^L_k$ denotes the computing frequency  (CPU cycles/s) of the terminal $k$ and $\beta$ (cycles/bit) denotes the task computation workload/intensity, which depends on the nature of the sensing task and can be estimated through task profilers \cite{MEC}, \cite{MECcof}. Terminal $k$'s power consumption under the local execution model is 
\begin{equation}
p_k^L=\eta(f^L_k)^3+||\mathbf{w}_{k}\mathbf{w}_{k}^H||^2,
\end{equation}
where $\eta$ is the power coefficient related to the chip architecture \cite{UAVMEC}.
\subsubsection{MEC Server Execution}
Compared to the local execution strategy, MEC severs are able to execute sensing computation tasks faster due to its enhanced computational capability, but it also introduces additional transmission latency while offloading sensing tasks. Specifically, the MEC server execution mode mainly consists of three steps: 1) Terminals offload their sensing tasks to their associated MEC servers. 2) The MEC server executes sensing task. 3) Terminals download the results from the MEC server. The execution latency $T_k^E$ for offloading task $\bm{C}_k$ to the MEC server $l$,  i.e., $a_{lk}=1, b_{lk}=0$, is expressed as
\begin{equation}\label{t2}
T_k^E=t_k^\text{UP}+t_k^\text{E}=\frac{{Z}_k}{R_{lk}}+\frac{\beta{Z}_k}{f^E_k}, ~\forall k\in \mathcal{K},
\end{equation}
where  $t_k^\text{UP}={{Z}_k}/{R_{lk}}$ and $t_k^\text{E}={\beta{Z}_k}/{f^E_k}$ denote the task offloading latency from terminal $k$ to MEC server $l$ and the sensing tasks execution latency on the MEC server, respectively. $f^E_k$ is the CPU frequency allocated to terminal $k$ by MEC server $l$. The corresponding power consumption of terminals primarily arises from ISAC signal transmission, which is expressed as
\begin{equation}
p_k^E=||\mathbf{w}_{k}\mathbf{w}_{k}^H||^2,\forall k\in \mathcal{K}.
\end{equation}

In practice, the computational capacity of each MEC server is usually limited, which induces the following constraint:
\begin{equation}
\sum_{k=1}^Ka_{lk}{f}_k^E \leq {G}_l , \forall l\in \mathcal{L},
\end{equation}
where ${G}_l$ is the computation capacity of MEC server $l$. 
\subsubsection{MCC Server Execution}
Although the computational capacity of the MEC server is higher than that of terminals, it is still limited, especially for a large-scale network. To alleviate the computation task offloading, the MEC server can further forward the received sensing computation tasks to the MCC server, i.e., $a_{lk}=0, b_{lk}=1$. In this case, the MCC server execution latency $T_k^C$ consists of three parts: the offloading latency
$t_k^\text{UP}$, the backhaul transmission latency $t_k^\text{Bh}$ from MEC server to MCC server, and the execution latency $t_k^\text{C}$ on the MCC server, which is expressed as
\begin{equation}\label{t3}
T_k^C=t_k^\text{UP}+t_k^\text{Bh}+t_k^\text{C}=\frac{{Z}_k}{R_{lk}}+\frac{{Z}_k}{r^{f}}+\frac{\beta{Z}_k}{f^C_k}, ~\forall k\in \mathcal{K},
\end{equation}
where $r^f$ is the transmission rate between MEC server and MCC server via optical fiber, and ${f^C_k}$ denotes the computation capability allocated to terminal $k$ by MCC server. The power consumption for terminals under MCC server execution mode is equivalent to that under the MEC server execution mode, i.e., $p_k^C = p_k^E$. 

\subsubsection{General Models} 
Consequently, terminal $k$'s total task execution latency is expressed as\footnote{The sensing task execution latency measures the freshness of the information carried by the echo signals, which is also referred to as the peak age of information (PAoI) \cite{AOI1}, \cite{AOI2}.}
\begin{equation}
\label{p1}
T_{k}^{total}=(1-~\sum\limits_{l\in \mathcal{L}}a_{lk}-~\sum\limits_{l\in \mathcal{L}}b_{lk})T_{k}^L+~\sum\limits_{l\in \mathcal{L}}{a_{lk}}T_{k}^E + ~\sum\limits_{l\in \mathcal{L}}{b_{lk}}T_{k}^C,
\end{equation}
and the total power consumption of terminal $k$ is 
\begin{equation}\label{p2}
p_{k}^{total}=(1-~\sum\limits_{l\in \mathcal{L}}a_{lk}-~\sum\limits_{l\in \mathcal{L}}b_{lk})p_{k}^L+~\sum\limits_{l\in \mathcal{L}}{a_{lk}}p_{k}^E + ~\sum\limits_{l\in \mathcal{L}}{b_{lk}}p_{k}^C.
\end{equation}

\subsection{Problem Formulation}
Based on the aforementioned analysis, it can be observed that reducing MTI through ISAC beamforming design effectively increases the uplink transmission rate $R_{lk}$, thereby reducing $T_{k}^E$ and $T_{k}^C$. \textcolor{black}{Furthermore, as shown in (\ref{p1}), different values of variables $\mathbf{a}$ and $\mathbf{b}$, which represent different offloading strategies, result in distinct total execution latencies $T_{k}^{total}$ for sensing tasks.} Therefore, the latency of sensing tasks is determined not only by the ISAC beamforming vectors but also by the task offloading strategy, while ISAC beamforming additionally impacts the SINR of echo signals. Consequently, the trade-off between sensing accuracy and execution latency must be addressed. In this paper, we focus on minimizing the total sensing task execution latency of all terminals by jointly optimizing the offloading decision variables and beamforming vectors under the constraints of sensing SINR, MEC servers' computation capacity, and terminals' power budget, which is formulated as follows
\begin{align} 
\label{prob29}
\min\limits _{\bm{a}_l,\bm{b}_l, \{\mathbf{w}_k\}} & ~\sum_{k=1}^{K}~T_{k}^{total}\\
s.t.~~~~
&\sum_{l=1}^La_{lk}+\sum_{l=1}^Lb_{lk}\leq 1, \forall k\in \mathcal{K}, \tag{\ref{prob29}a}\\
& {a_{lk}},{b_{lk}} \in \left\{ {0,1} \right\},~\forall l\in \mathcal{L},k\in \mathcal{K},\tag{\ref{prob29}b}\\
& \sum_{k=1}^Ka_{lk} {f}^E_k \leq {G}_l , \forall l\in \mathcal{L}, \tag{\ref{prob29}c}\\
& p_{k}^{total} \leq P_{th},\forall k\in \mathcal{K}, \tag{\ref{prob29}d}\\
&{\gamma}^r_{k} \geq \Gamma_{r},\forall k\in \mathcal{K},\tag{\ref{prob29}e}
\end{align}
where $P_{th}$ is terminal's maximum power available for sensing, communication, and computing, and $\Gamma_{r}$ is the sensing SINR threshold. The problem ({\ref{prob29}}) is challenging to solve due to the presence of integer variables $\bm{a}_l, \bm{b}_l$, non-convex objective function and non-convex constraint ({\ref{prob29}}e). Moreover, it can be observed that in the objective function, there exists coupling between the offloading decision variables and the beamforming vectors, which makes it harder to solve the problem (\ref{prob29}).

\section{Joint ISAC Beamforming and Computation Offloading Design}
As the network scale increases, the centralized algorithms may incur excessive computational complexity, imposing a significant computational burden on the network. Therefore, in this section, we propose a distributed alternating optimization algorithm to solve problem (\ref{prob29}), where all BSs participate in the optimization. Specifically, we decompose problem (\ref{prob29}) into two sub-problems. For given $\mathbf{w}_k$'s, the offloading decision variables $\bm{a}_l, \bm{b}_l$ are first optimized based on the distributed  ADMM framework \cite{XieTSP}. For given offloading decision variables $\bm{a}_l, \bm{b}_l$, we then optimize the beamforming vectors based on the distributed WMMSE algorithm and FP technique. The convergence of the alternating optimization algorithm will be validated by numerical results in Section IV. \textcolor{black}{This distributed approach can significantly enhance system scalability, as adding more nodes does not place \textcolor{black}{heavier} load on the central server. Additionally, by distributing the computational load across multiple MEC servers, the overall resource can be fully taken advantage of.}
\subsection{Computation Offloading Optimization}
In this subsection, we focus on optimizing the offloading decision variables $\bm{a}_k, \bm{b}_k$ with given $\mathbf{w}_k$'s. 

To handle the binary variables, we relax $\bm{a}_k, \bm{b}_k$ into continuous variables $0\leq{a_{lk}}\leq 1,0\leq {b_{lk}}\leq 1$. The computation offloading sub-problem is then relaxed as
\begin{align} 
\label{prob34}
\min\limits _{\bm{a}_l,\bm{b}_l} &~\sum_{k=1}^{K}~T_{k}^{total}\\
s.t.~
& 0\leq{a_{lk}},{b_{lk}}\leq 1,~ \forall l\in \mathcal{L},k\in \mathcal{K},\tag{\ref{prob34}a},\\
& (\ref{prob29}\text{a}),(\ref{prob29}\text{c}) ,(\ref{prob29}\text{d}),\notag
\end{align}

In order to solve the problem in a distributed manner, we express the objective function in a decomposable form with respect to different BSs, which is
\begin{align} 
&\sum_{k=1}^{K}~T_{k}^{total} \notag\\
&\quad=\sum_{l=1}^{L}~\sum_{k=1}^{K}(\frac{1}{L}T_{k}^L+a_{lk}(T_k^E-T_k^L)+b_{lk}(T_k^C-T_k^L)) \notag\\
&\quad\triangleq\sum_{l=1}^{L}~f_{l}(\bm{a}_l,\bm{b}_l)
\end{align}

To enable all BSs to independently solve the problem, we introduce the local copies of ${a}_{lk}$ and  ${b}_{lk}$ for BS $l$, which are denoted as $\omega_{lk}$ and $\varpi_{lk}$, respectively. The feasible sets for ${\bm{\omega}_l,\bm{\varpi}_l}$ and ${a_{lk},b_{lk}}$ for all $l \in \mathcal{L}$ and $k \in \mathcal{K}$ are denoted by $\{\mathcal{U}_l\}_{l\in \mathcal{L}}$ and $\mathcal{V}$, respectively, which are expressed as
\begin{equation}
\label{fs1}
{\mathcal{U}_l} = \left\{ {\left. \begin{array}{l}
	{\bm{\omega}_l}\\
	{\bm{\varpi}_l}
	\end{array} \right|\begin{array}{*{20}{c}}
	0\leq{\omega_{lk}},{\varpi_{lk}}\leq 1,~ \forall k\in \mathcal{K}\\
	\sum_{k=1}^K\omega_{lk}{f}^E_k \leq {G}_l \\
	\end{array}} \right\},
\end{equation}
\begin{equation}
\label{fs2}
{\mathcal{V}} = \left\{ {\left. \begin{array}{l}
	{a_{lk}}\\
	{b_{lk}}
	\end{array} \right|\begin{array}{*{20}{c}}
	\sum_{l=1}^La_{lk}+\sum_{l=1}^Lb_{lk}\leq 1, \forall k\in \mathcal{K}\\
	p_{k}^{total} \leqslant P_{th}, \forall k\in \mathcal{K} \\
	\end{array}} \right\}.
\end{equation}
The consensus counterpart of problem (\ref{prob34}) is formulated as
\begin{align} 
\label{prob35}
\min\limits _{\{\mathcal{U}_l\}_{l\in \mathcal{L}},\mathcal{V}} &~\sum_{l=1}^{L}~f_{l}(\bm{\omega}_l,\bm{\varpi}_l)\\
s.t.~~~
& \omega_{lk}={a_{lk}},~ \forall l\in \mathcal{L},k\in \mathcal{K},\tag{\ref{prob35}a}\\
& \varpi_{lk}={b_{lk}},~ \forall l\in \mathcal{L},k\in \mathcal{K},\tag{\ref{prob35}b}
\end{align}

Under the ADMM framework, the augmented Lagrangian function of problem (\ref{prob35}) is formulated as
\begin{equation}
\begin{aligned}
\label{lg1}
&\mathcal{L}\left( \{\mathcal{U}_l\}_{l\in \mathcal{L}},\mathcal{V},\bm{\hat{\phi}},\bm{\hat{\varphi}} \right)\\ 
&= \sum_{l=1}^{L}~f_{l}(\bm{\omega}_l,\bm{\varpi}_l) + \sum\limits_{l \in \mathcal{L}} {\sum\limits_{\substack{k \in \mathcal {K}}} {{\phi _{lk}\left( {\omega_{lk} - {a_{lk}}} \right)}} } \\
&+\sum\limits_{l \in \mathcal{L}} {\sum\limits_{\substack{k \in \mathcal {K}}} {{{\varphi _{lk}\left( {\varpi_{lk} - {b_{lk}}} \right)}}} }+{\frac{\rho }{2}\sum\limits_{l \in \mathcal{L}} {\sum\limits_{\substack{k \in \mathcal {K}}} {{{\left\| {\omega_{lk} - {a_{lk}}} \right\|}^2}} } }\\
&+{\frac{\rho }{2}\sum\limits_{l \in \mathcal{L}} {\sum\limits_{\substack{k \in \mathcal {K}}} {{{\left\| {\varpi_{lk} - {b_{lk}}} \right\|}^2}} } },
\end{aligned}
\end{equation}
where $\bm{\hat{\phi}}=\left\{\phi _{i,k}\right\}_{\forall l\in \mathcal{L},k\in \mathcal{K}}$ and $\bm{\hat{\varphi}}=\left\{\varphi_{i,k}\right\}_{\forall l\in \mathcal{L},k\in \mathcal{K}}$ are the Lagrange multipliers associated with the consensus constraints (\ref{prob35}a) and (\ref{prob35}b), and $\rho$ is the penalty factor. Through combining the linear and quadratic terms of the consensus constraints, (\ref{lg1}) is rewritten as
\begin{equation}
\begin{aligned}
\label{lg2}
&\mathcal{L}\left( \{\mathcal{U}_l\}_{l\in \mathcal{L}},\mathcal{V},\bm{\hat{\phi}},\bm{\hat{\varphi}} \right)\\ 
&= \sum_{l=1}^{L}~f_{l}(\bm{\omega}_l,\bm{\varpi}_l)+{\frac{\rho }{2}\sum\limits_{l \in \mathcal{L}} {\sum\limits_{\substack{k \in \mathcal {K}}} {{{\left\| {\omega_{lk} - {a_{lk}+\hat{\phi}_{lk}}} \right\|}^2}} } }\\
&+{\frac{\rho }{2}\sum\limits_{l \in \mathcal{L}} {\sum\limits_{\substack{k \in \mathcal {K}}} {{{\left\| {\varpi_{lk}^m - {b_{lk}+\hat{\varphi}_{lk}}} \right\|}^2}} } },
\end{aligned}
\end{equation}
where $\hat{\phi}_{lk}=\frac{\phi_{lk}}{\rho}$ and $\hat{\varphi}_{lk}=\frac{\varphi_{lk}}{\rho}$. The dual problem of problem (\ref{prob35}) is
\begin{equation}
\label{prob36}
\mathop {\max }\limits_{{\bm{\hat{\phi}}, \bm{\hat{\varphi}}}} ~\mathcal{L}_{d}\left( {\bm{\hat{\phi}}, \bm{\hat{\varphi}}} \right),
\end{equation}
where $\mathcal{L}_{d}$ is the dual function given by
\begin{equation}
\label{prob37}
\mathcal{L}_{d}\left( {\bm{\hat{\phi}}, \bm{\hat{\varphi}}} \right)=\mathop {\min }\limits_{ \{\mathcal{U}_l\}_{l\in \mathcal{L}},\mathcal{V}} ~\mathcal{L}\left({{  \{\mathcal{U}_l\}_{l\in \mathcal{L}},\mathcal{V}}}, {\bm{\hat{\phi}}, \bm{\hat{\varphi}}} \right).
\end{equation}

By iteratively updating $\{\mathcal{U}_l\}_{l\in \mathcal{L}}$, $\mathcal{V}$ and $\{\bm{\hat{\phi}}, \bm{\hat{\varphi}}\}$ via solving (\ref{prob36}) and (\ref{prob37}), the orignal problem (\ref{prob35}) can be solved. By denoting $N_{it}$ as the iteration number, the detailed ADMM algorithm for updating the three variables consists of the following steps.

\textit {a) Update $\{\mathcal{U}_l\}_{l\in \mathcal{L}}$}: With fixed $\mathcal{V}^{N_{it}}$ and $\{\bm{\hat{\phi}}^{N_{it}}, \bm{\hat{\varphi}}^{N_{it}}\}$, the optimal local variables $\{\mathcal{U}_l\}_{l\in \mathcal{L}}$ are expressed as
\begin{equation}
\label{u1}
{{\{\mathcal{U}_l\}^{\left( {N_{it}+1} \right)}_{l\in \mathcal{L}}}} = \mathop {\arg \min }\limits_{\{\mathcal{U}_l\}_{l\in \mathcal{L}}} {L }\left( {{  \{\mathcal{U}_l\}_{l\in \mathcal{L}},\mathcal{V}^{N_{it}},{\bm{\hat{\phi}}^{N_{it}}, \bm{\hat{\varphi}}^{N_{it}}}}} \right).
\end{equation}

Problem (\ref{u1}) can be separated into $l$ sub-problems, which are solved by each BS independently. For BS $l$, the optimization subproblem is given by
\begin{equation}
\begin{aligned}
\label{u2}
\min_{\bm{\omega}_l,\bm{\varpi}_l} ~&f_{l}(\bm{\omega}_l,\bm{\varpi}_l)+{\frac{\rho }{2}{\sum\limits_{\substack{k \in \mathcal {K}}} {{{\left\| {\omega_{lk} - {a_{lk}+\hat{\phi}_{lk}}} \right\|}^2}} } }\\
&+{\frac{\rho }{2}{\sum\limits_{\substack{k \in \mathcal {K}}} {{{\left\| {\varpi_{lk}^m - {b_{lk}+\hat{\varphi}_{lk}}} \right\|}^2}} } },\\
s.t.\quad 
&\bm{\omega}_l,\bm{\varpi}_l\in {\mathcal{U}_l},
\end{aligned}
\end{equation}
which is a convex problem and can be solved by the interior-point algorithm \cite{2004Convex}. 

\textit {b) Update $\mathcal{V}$}: With fixed $\{\mathcal{U}_l\}^{(N_{it}+1)}_{l\in \mathcal{L}}$ and $\{\bm{\hat{\phi}}^{N_{it}}, \bm{\hat{\varphi}}^{N_{it}}\}$, the optimal global variables $\mathcal{V}$ are given as
\begin{equation}
\label{u3}
{{\mathcal{V}^{\left( {N_{it}+1} \right)}}} = \mathop {\arg \min }\limits_{\mathcal{V}} {L }\left( {{ {\bm{\hat{\phi}}^{N_{it}}, \bm{\hat{\varphi}}^{N_{it}}}. \{\mathcal{U}_l\}^{(N_{it}+1)}_{l\in \mathcal{L}},\mathcal{V}}} \right),
\end{equation}
and the corresponding optimization problem is formulated as
\begin{align}
\label{u4}
\min_{\bm{a}_l,\bm{b}_l} ~&{\frac{\rho }{2}\sum\limits_{l \in \mathcal{L}} {\sum\limits_{\substack{k \in \mathcal {K}}} {{{\left\| {\omega_{lk} - {a_{lk}+\hat{\phi}_{lk}}} \right\|}^2}} } } \notag\\&+{\frac{\rho }{2}\sum\limits_{l \in \mathcal{L}} {\sum\limits_{\substack{k \in \mathcal {K}}} {{{\left\| {\varpi_{lk}^m - {b_{lk}+\hat{\varphi}_{lk}}} \right\|}^2}} } }\\
s.t.&\quad 
\sum_{l=1}^La_{lk}+\sum_{l=1}^Lb_{lk}\leq 1, \forall k\in \mathcal{K}\tag{\ref{u4}a}\\
&\quad p_{k}^{total} \leq P_{th}, \forall k\in \mathcal{K} ,\tag{\ref{u4}b}
\end{align}

By introducing the Lagrange multipliers $\{\chi_k\}_{k\in \mathcal{K}}$ and $\{\psi_k\}_{k\in \mathcal{K}}$ to constraints (\ref{u4}a) and (\ref{u4}b), the Lagrangian function of problem (\ref{u4}) is derived as
\begin{equation} \small
\begin{aligned}
\label{lg3}
&\mathcal{L}\left(\bm{a}_l,\bm{b}_l,\{\chi_k\}_{k\in \mathcal{K}},\{\psi_k\}_{\in \mathcal{K}} \right)={\frac{\rho }{2}\sum\limits_{l \in \mathcal{L}} {\sum\limits_{\substack{k \in \mathcal {K}}} {{{\left\| {\omega_{lk} - {a_{lk}+\hat{\phi}_{lk}}} \right\|}^2}} } } \notag\\&+{\frac{\rho }{2}\sum\limits_{l \in \mathcal{L}} {\sum\limits_{\substack{k \in \mathcal {K}}} {{{\left\| {\varpi_{lk}^m - {b_{lk}+\hat{\varphi}_{lk}}} \right\|}^2}} } }+\sum_{k=1}^K\chi_k\left(1-\sum_{l=1}^La_{lk}-\sum_{l=1}^Lb_{lk}\right)\\
&+\sum_{k=1}^K\psi_k\left(P_{th}-p_{k}^{total} \right),
\end{aligned}
\end{equation}

Based on the Karush-Kuhn-Tucker (KKT) conditions \cite{2004Convex}, the optimal closed-form solutions of ${a}_{lk}$ and ${b}_{lk}$ are derived as follows.
	\begin{equation}
	\label{a1}
{a}^{(N_{it}+1)}_{lk}(\chi_k,\psi_k)=\frac{1}{\rho}(-\chi_k-\psi_k(p_k^E-p_k^L))+\omega^{(N_{it}+1)}_{lk}+\hat{\phi}^{N_{it}}_{lk}, 
\end{equation}
\begin{equation}
\label{b1}
{b}^{(N_{it}+1)}_{lk}(\chi_k,\psi_k)=\frac{1}{\rho}(-\chi_k-\psi_k(p_k^C-p_k^L))+\varpi_{lk}^{(N_{it}+1)}+\hat{\varphi}^{N_{it}}_{lk}. 
\end{equation}

\textcolor{black}{It can be readily seen that ${a}^{(N_{it}+1)}_{lk}$ and ${b}^{(N_{it}+1)}_{lk}$ are both linear functions with respect to  $\chi_k$ and $\psi_k$, and according to (\ref{p2}), $p_{k}^{total}(\chi_k,\psi_k)$ is also the linear function with respect to  $\chi_k$ and $\psi_k$. By substituting these linear functions of $\{\chi_k\}_{k\in \mathcal{K}}$ and $\{\psi_k\}_{k\in \mathcal{K}}$ into complementary slackness conditions \cite{2004Convex}, we have}
\begin{align}
&\chi_k\left(\sum_{l=1}^La_{lk}(\chi_k,\psi_k)+\sum_{l=1}^Lb_{lk}(\chi_k,\psi_k)- 1\right)=0, \forall k\in \mathcal{K}\\
&\psi_k\left(p_{k}^{total}(\chi_k,\psi_k) -P_{th}\right)=0, \forall k\in \mathcal{K}.
\end{align}
where the optimal $\{\chi_k\}_{k\in \mathcal{K}}$ and $\{\psi_k\}_{k\in \mathcal{K}}$ can be obtained by subgradient based methods, e.g., the ellipsoid method in \cite{Feng2018}.

\textit {c) Update $\{\bm{\hat{\phi}}, \bm{\hat{\varphi}}\}$}: With fixed $\mathcal{V}^{N_{it}}$ and $\{\mathcal{U}_l\}_{l\in \mathcal{L}}$, the  optimal Lagrange multipliers variables $\{\bm{\hat{\phi}}, \bm{\hat{\varphi}}\}$ are expressed as
\begin{equation}
\label{u5}
\{{\bm{\hat{\phi}}, \bm{\hat{\varphi}}}\}^{(N_{it}+1)} = \mathop {\arg \min }\limits_{{\bm{\hat{\phi}}, \bm{\hat{\varphi}}}} {L }\left( {{ {\bm{\hat{\phi}}, \bm{\hat{\varphi}}}. \{\mathcal{U}_l\}^{\left( {N_{it}+1} \right)}_{l\in \mathcal{L}},\mathcal{V}^{(N_{it}+1)}}} \right).
\end{equation}

By applying the gradient descent method in the ADMM framework, $\{\bm{\hat{\phi}}, \bm{\hat{\varphi}}\}$ can be updated by
\begin{equation}
\label{a3}
{\hat{\phi}}^{(N_{it}+1)}_{(lk)}={\hat{\phi}^{(N_{it})}_{{lk}}+\upsilon(\omega^{(N_{it}+1)}_{lk} - a^{(N_{it}+1)}_{lk})}, 
\end{equation}

\begin{equation}
\label{b3}
{\hat{\varphi}}_{(lk)}^{(N_{it}+1)}={\hat{\varphi}^{(N_{it})}_{lk}+\upsilon(\varpi^{(N_{it}+1)}_{lk} - {b^{(N_{it}+1)}_{lk}}}), 
\end{equation}
where $\upsilon$ is the iteration step size. 
  
\textbf{Summary:} To make it clearer,  we summarize the detailed process of the ADMM framework for solving problem (\ref{prob34}) in Algorithm 1. 
To recover the binary variables based on the optimized continuous variables, we utilize the continuous relaxation and inflation based approach in \cite{tsp3}. 
\begin{algorithm}[t] 
	\caption{The Algorithm based on ADMM framework for solving problem (\ref{prob34})  }
	\begin{algorithmic}[1]
		\renewcommand{\algorithmicrequire}{\textbf{Input:}}
		\renewcommand{\algorithmicensure}{\textbf{Output:}}
		\Require Beamforming vectors $\mathbf{w}_k, \forall k$, maximum iteration number $iter_{max}$, the Lagrange multipliers ${\bm{\hat{\phi}}, \bm{\hat{\varphi}}}$ and the convergence threshold $\epsilon>0$,
		
		\State {Initialize $\{\bm{a}_l, \bm{b}_l\}_{l\in\mathcal{L}}$ as a feasible solution.}
	
		\Repeat
		
	    	\State {Each BS updates local offloading decision variables $\{\bm{\omega}_l,\bm{\varpi}_l\}$. }
		
		\State {Update the global offloading decision variables $\bm{a}$ and $\bm{b}$ based on (\ref{a1}) and (\ref{b1}) .}
		\State {Update the Lagrange multipliers variables $\{\bm{\hat{\phi}}, \bm{\hat{\varphi}}\}$ based on (\ref{a3}) and (\ref{b3}). }	
		\Until {$|\bm{a}^{(N_{it}+1)}-\bm{a}^{N_{it}}|\leq\epsilon_0 $,$|\bm{b}^{(N_{it}+1)}-\bm{b}^{N_{it}}|\leq\epsilon_0 $.}
		\Ensure The optimal offloading decision variables $\bm{a}$ and $\bm{b}$.
	\end{algorithmic}
\end{algorithm}

\textit{\textbf{Complexity Analysis:}} The complexity of Algorithm 1 consists of updating block variables $\{\mathcal{U}_l\}_{l\in \mathcal{L}}$, $\mathcal{V}$ and $\{\bm{\hat{\phi}}, \bm{\hat{\varphi}}\}$. For $\{\mathcal{U}_l\}_{l\in \mathcal{L}}$, the complexity of solving problem (\ref{u2}) is $\mathcal{O}\left(K^{3}\right)$. The complexity for updating the global optimization variable $\mathcal{V}$ is $\mathcal{O}\left(KL\right)$, and the complexity for updating the Lagrange multiplier $\{\bm{\hat{\phi}}, \bm{\hat{\varphi}}\}$ is $\mathcal{O}\left(KL\right)$. Therefore, in each iteration, the overall computational complexity is $\mathcal{O}\left(KL+K^{3}\right)$. 

\subsection{ISAC Beamforming Optimization}

In this subsection, we tackle the optimization of beamforming vectors $\mathbf{w}_k$'s with given offloading decision variables. According to (\ref{t1}), (\ref{t2}), and (\ref{t3}), when $\bm{a}_l$'s, $\bm{b}_l's$ are fixed, the objective function $T_{k}^{total}$ depends solely on $R_{lk}$. We define the set of terminals offloading sensing tasks to BS $l$, i.e., ${a}_{lk}=1$ or ${b}_{lk}=1$, as $\mathcal{B}_l$. By omitting constant terms, the beamforming optimization sub-problem can be expressed as

\begin{align} 
\label{prob30}
\min \limits _{\{\mathbf{w}_k\}} & \sum_{l\in \mathcal{L}} \sum_{k\in\mathcal{B}_l}~\frac{{Z}_k}{R_{lk}}\\
s.t.~
& p_{k}^{total} \leq P_{th}, \forall k\in \mathcal{K}, \tag{\ref{prob30}a}\\
&{\gamma}^r_{k} \geq \Gamma_k,\forall k\in \mathcal{K}, \tag{\ref{prob30}b}
\end{align}
which is a sum-of-ratio problem. By employing the quadratic transform for multiple-ratio FP \cite{fp4} and  introducing auxiliary variable $\bm{c}=\{c_{k}\}$, we further rewrite the problem (\ref{prob30}) as
\begin{align} 
\label{prob31}
\min \limits _{\{\mathbf{w}_k\},\{c_k\}} & \sum_{l\in \mathcal{K}} \sum_{k\in\mathcal{B}_l}~ 2\sqrt{Z_k}c_{k}-c^2_{k}{R_{lk}}\\
s.t.~
& p_{k}^{total} \leq P_{th}, ~~\forall k\in \mathcal{L},\tag{\ref{prob31}a}\\
&{\gamma}^r_{k} \geq \Gamma_k,~~\forall k\in \mathcal{K}.\tag{\ref{prob31}b}
\end{align}

Problem (\ref{prob31}) can be solved by iteratively optimizing $c_k$ and $\mathbf{w}_k$. When $\mathbf{w}_k$ is fixed, the optimal $c_k$ is given by
\begin{equation}
\label{up1}
c^*_k=\frac{\sqrt{Z_k}}{R_{lk}}, \forall k,
\end{equation}

To facilitate the subsequent derivation, we introduce the auxiliary weight variables $\delta_k$. For $k\in\mathcal{B}_l$, $\delta_k=1$;  otherwise, $\delta_k=0$. Thus, the problem (\ref{prob31}) with fixed $c_k$'s can be reformulated as
\begin{align} 
\label{prob32}
\max \limits _{\{\mathbf{w}_k\}} & \sum_{l\in \mathcal{L}} \sum_{k\in\mathcal{K}}~ c^2_{k}\delta_k{R_{lk}}\\
s.t.~
& (\ref{prob31}\text{a}),(\ref{prob31}\text{b}),\notag
\end{align}

To address the non-convex problem (\ref{prob32}), we employ an iterative WMMSE based method\cite{sqjtsp}. By assuming that the BS $l$ utilizes the linear receiver $\mathbf{u}_k$ for user $k$'s signal detection, the mean square error (MSE)  can be expressed as

\begin{equation}
\begin{aligned}
\label{mse}
E_k&=\mathbb{E}\left[\left(\mathbf{u}_k \mathbf{y}_l-\mathbf{s}_k\right)\left(\mathbf{u}_k \mathbf{y}_l-\mathbf{s}_k\right)^{\mathrm{H}}\right] \\
&=\mathbf{u}_k\mathbf{F}_k\mathbf{u}^H_k-\mathbf{G}_k-\mathbf{G}^H_k+{\sigma_b}^2\mathbf{u}_k\mathbf{u}_k^H
\end{aligned}
\end{equation}
where
\begin{equation}
\mathbf{F}_k=\sum_{i=1}^{K}\mathbf{H}^H_{li}\mathbf{w}_{i}\mathbf{w}^H_{i}\mathbf{H}_{li},
\end{equation}
and 
\begin{equation}
\mathbf{G}_k=\mathbf{u}_k\mathbf{H}^H_{lk}\mathbf{w}_{k},
\end{equation}

With given transmit beamformers $\mathbf{w}_k$, the optimal $\mathbf{u}_k$ can be obtained by the well-known MMSE receiver:
\begin{equation}
\label{mmse}
\mathbf{u}^{opt}_k = \mathbf{w}^H_{k}\mathbf{H}_{lk}({\sigma_b}^2\mathbf{I}_M+\mathbf{F}_k)^{-1}.
\end{equation}

By substituting (\ref{mmse}) into (\ref{mse}),  the MSE can be rewritten as
\begin{equation}
E_k = 1-\mathbf{w}^H_{k}\mathbf{H}_{lk}({\sigma_b}^2\mathbf{I}_M+\mathbf{F}_k)^{-1}\mathbf{H}^H_{lk}\mathbf{w}_{k}.
\end{equation}

For ease of expression, we  denote $\{\mathbf{u}_k\}_{\forall k\in\mathcal{K}}$ as $\mathbf{u}$. Next, by introducing auxiliary weight vectors $\mathbf{v}=\{V_k\}_{\forall k\in\mathcal{K}}$, we have the following relationship between $R_{lk}$ and $E_k$.

\begin{lemma}
The rate expression $R_{lk}$ can be equivalently represented as
\begin{equation}
\begin{aligned}
\label{r1u}
R_{lk}&=B(\log(V_k)-V_kE_k+1)\\&=\widehat{R}_{lk}(\mathbf{u}_k,V_k,\mathbf{w}_k),
\end{aligned}
\end{equation}
and the optimal $V_k$ is given by
\begin{equation}
\label{MMSE2}
V^{opt}_k=(1-\mathbf{w}^H_{k}\mathbf{H}_{lk}({\sigma_b}^2\mathbf{I}_M+\mathbf{F}_k)^{-1}\mathbf{H}^H_{lk}\mathbf{w}_{k})^{-1}.
\end{equation}
\end{lemma}
\begin{IEEEproof}
 Please see Appendix A.
\end{IEEEproof}

\textcolor{black}{For given $\mathbf{u}_k$ and $V_k$, by omitting the constant terms, (\ref{r1u})} can be reformulated as
\begin{equation}
\begin{aligned}
\widehat{R}_{lk}(\mathbf{w}_k)&=\text{tr}(V_k\mathbf{u}_k\mathbf{H}^H_{lk}\mathbf{w}_{k})+\text{tr}(V_k\mathbf{w}^H_{k}\mathbf{H}_{lk}\mathbf{u}_k^H)\\&-\sum_{i=1}^{K}\text{tr}(V_k\mathbf{u}_k\mathbf{H}^H_{li}\mathbf{w}_{i}\mathbf{w}^H_{i}\mathbf{H}_{li}\mathbf{u}^H_k).
\end{aligned}
\end{equation}
and the optimization problem of $\mathbf{w}_k$ is 
\begin{align} 
\label{prob22}
\max \limits _{\{\mathbf{w}_k\}} & \sum_{l\in \mathcal{L}} \sum_{k\in\mathcal{K}}~c_{k}^2\delta_k\widehat{R}_{lk}\\
s.t.~
& (\ref{prob31}\text{a}),(\ref{prob31}\text{b}),\notag
\end{align}

To solve the problem (\ref{prob22}) in a distributed manner, we decompose the objective function  with respect to $\mathbf{w}_k$ as
\begin{equation}
\begin{aligned}
{f}_{k}(\mathbf{w}_k)&=-c^2_{k}\delta_k \text{tr}(V_k\mathbf{u}_k\mathbf{H}^H_{li}\mathbf{w}_{k}\mathbf{w}^H_{k}\mathbf{H}_{li}\mathbf{u}^H_k) \\&-\sum_{l\in L} \sum_{i\in\mathcal{K},i\neq k}c^2_{i}\delta_i\text{tr}(V_i\mathbf{u}_i\mathbf{H}^H_{lk}\mathbf{w}_{k}\mathbf{w}^H_{k}\mathbf{H}_{lk}\mathbf{u}^H_i)\\&+c^2_{k}\delta_k\text{tr}(V_k\mathbf{u}_k\mathbf{H}^H_{lk}\mathbf{w}_{k})+c^2_{k}\delta_k\text{tr}(\mathbf{w}^H_{k}\mathbf{H}_{lk}\mathbf{u}_k^HV_k)
\end{aligned}
\end{equation}

 By introducing auxiliary variables $\bm{\Upsilon}=\{\Upsilon_{kj}\}_{\forall k,j}$, constraint (\ref{prob31}\text{b}) can be rewritten as
\begin{align}
&\frac{\xi_k^2\text{tr}(\mathbf{A}({\theta}_k)\mathbf{w}_{k}\mathbf{w}^H_{k}\mathbf{A}({\theta}_k)^H)} {{\sum_{j=1,j\neq k}^K\Upsilon_{kj}}+\sigma_k^2}\geq \Gamma_k, \forall k,\label{dc1}\\
&\text{tr}({{\mathbf{H}}_{kj}^{IH}}\mathbf{w}_{j}\mathbf{w}^H_{j}{\mathbf{H}}^{I}_{kj})\leq\Upsilon_{kj}, \forall k\neq j.\label{dc2}
\end{align}

We can observe that constraint (\ref{dc1}) remains non-convex due to the quadratic terms. Therefore, we employ the SCA technique \cite{sca} to convexify it. Specifically, the constraint (\ref{dc1}) at a given point $\widetilde{\mathbf{w}}_{k}$ can be approximated as 
\begin{equation}
\begin{aligned}
\label{prob3} 
{ 2\text{tr}(\widetilde{\mathbf{w}}^H_{k}\mathbf{A}({\theta}_k)^H\mathbf{A}({\theta}_k)\mathbf{w}_{k})-\text{tr}(\widetilde{\mathbf{w}}^H_{k}\mathbf{A}({\theta}_k)^H\mathbf{A}({\theta}_k)\widetilde{\mathbf{w}_{k}})} \\
\geq \frac{\Gamma_{k}}{\xi_k^2}({\sum_{j=1,j\neq k}^K\Upsilon_{kj}}+\sigma_k^2),~~\forall k\in\mathcal{K}.
\end{aligned}
\end{equation}

Based on the aforementioned transformation, the problem (\ref{prob32}) with a given $\bm{\Upsilon}$ can be decomposed into $K$ sub-problems w.r.t ${\mathbf{w}}_{k}$:
\begin{align} 
\label{prob23}
{\mathcal{P}}_k^{\mathrm{sub}}({\mathbf{w}}_{k}):\max \limits _{\mathbf{w}_k} &  ~~~{f}_{k}(\mathbf{w}_k)\\
s.t.~&\text{tr}({{\mathbf{H}}_{jk}^{IH}}\mathbf{w}_{k}\mathbf{w}^H_{k}{\mathbf{H}}^{I}_{jk})\leq\Upsilon_{jk}, \forall k,j\in\mathcal{K},j\neq k  \tag{\ref{prob23}a}\\
& \text{(\ref{prob31}\text{a})},\text{(\ref{prob3})}, \notag
\end{align}
\noindent which is a standard convex problem and can be solved via the interior point method \cite{2004Convex}. After decomposition, BS $l$ can distributively solve the beamforming optimization sub-problems corresponding to its connected terminals. After updating $\{\mathbf{w}_{k}\}^K_{k=1}$, The update of $\bm{\Upsilon}$
can be achieved by solving the following optimization problem.
\begin{align} 
\label{prob24}
\qquad\text{Find}:&~~{\bm{\Upsilon}} \\
s.t.~&\text{tr}({{\mathbf{H}}_{jk}^{IH}}\mathbf{w}_{k}\mathbf{w}^H_{k}{\mathbf{H}}^{I}_{jk})\leq\Upsilon_{jk}, \forall k,j\in\mathcal{K},j\neq k, \tag{\ref{prob24}a}\\
& \text{(\ref{prob3})}, \notag
\end{align}
which can be solved via the CVX toolbox. To reduce the computational complexity, we adopt the updating criterion for $\Upsilon_{jk}$ in the following lemma.

\begin{lemma}
A feasible solution to problem (\ref{prob24}) is
{\begin{equation}
	\label{gamaup}
	\Upsilon^*_{jk}={\text{tr}}({{\mathbf{H}}_{jk}^{IH}}\mathbf{w}_{k}\mathbf{w}^H_{k}{\mathbf{H}}^{I}_{jk}), \forall k,j\in\mathcal{K},j\neq k.
	\end{equation}}
\end{lemma}

\begin{IEEEproof}
	Please see Appendix B.
\end{IEEEproof}
\renewcommand{\algorithmicrequire}{\textbf{Input:}}
\renewcommand{\algorithmicensure}{\textbf{Output:}}
\begin{algorithm}[t] 
	\caption{Alternating Optimization Algorithm for Solving (\ref{prob30}) }
	\begin{algorithmic}[1] 
		\Require Offloading decision variables $\bm{a}_l, \bm{b}_l$, initial iteration number $N_{it}=0$, maximum iteration number $N_{itmax}$, $\Gamma_k,P_{th}$,  convergence thresholds $\epsilon>0$,
		
		\State {Initialize  $\bm{c}$ , $\bm{\Upsilon}$ and $\{\mathbf{w}_k\}_{\forall k}$ to a feasible value.}
		\Repeat
		\State {Update $\mathbf{u}$ according to (\ref{mmse}). }
		\State {Update  $\mathbf{v}$ according to (\ref{MMSE2}).}
		\State {Each BS updates the associated ISAC beamforming vectors $\{\mathbf{w}_{k}\}_{k\in \mathcal{B}_l}$ by solving problem (\ref{prob23}).}
		\State {Update  $\bm{\Upsilon}$ by equation (\ref{gamaup}).}
		\State {Update  $\bm{c}$ by equation (\ref{up1}).}	  	    
		\State {Update $N_{it} = N_{it} + 1$.}
		\Until {$N_{it} \geq N_{itmax}$ or the  increase of (\ref{prob30}) is smaller than $\epsilon$.}

		\Ensure the beamforming vectors $\{\mathbf{w}_k\}_{\forall k}$.
	\end{algorithmic}
\end{algorithm}

By employing the BCD technique, we iteratively update varibles $\mathbf{u}$, $\mathbf{v}$, $\{\mathbf{w}_k\}_{\forall k\in\mathcal{K}}$, $\bm{\Upsilon}$ and $\bm{c}$, thereby solving the original sum of ratios problem
(\ref{prob30}). The detailed algorithm is summarized in Algorithm 2.

\textit{\textbf{Complexity Analysis:}} The computational complexity of Algorithm 2 comprises four main components: updating block optimization variables $\mathbf{u}$, $\mathbf{v}$, $\{\mathbf{w}_k\}$, and $\bm{\Upsilon}$. The complexity of updating $\mathbf{u}$ mainly comes from matrix inversion and matrix multiplication operations, requiring  $\mathcal{O}(MN+M^2N+M^3+M^2)$. Similarly, $\mathbf{v}$ and $\bm{\Upsilon}$ each require $\mathcal{O}(N^2)$ and $\mathcal{O}(K^2N^3)$, respectively. For updating $\{\mathbf{w}_k\}$, the complexity primarily resides in problem (\ref{prob23}), which can be solved by the primal-dual interior-point algorithm \cite{2004Convex}.  Its worst-case computational complexity can be analyzed by leveraging the analytical treatment in \cite{2001Convex}, which is $\mathcal{O}\left(N^{3}+K^{1.5}N^2\right)$. Consequently, by omitting lower order terms, the overall complexity of Algorithm 2 is $\mathcal{O}(K^{2}N^3+M^3+M^2N)$.

\textbf{Remark 2.} \textcolor{black}{\textit{In practical network deployments, the implementation of Algorithms 1 and 2 necessitates an exchange of information between base stations, MEC servers, the MCC server, and ISAC terminals. Specifically, base stations equipped with MEC servers initially collect terminals' channel state information and data offloading demands, and then forward this information to the MCC server. The MCC server, following the decomposition steps outlined in Algorithms 1 and 2, breaks down the optimization tasks for the beamforming vectors and offloading decisions, which are then distributed to various MEC servers. Upon completing the optimization of each subtask, the MEC servers send the locally optimized beamforming vectors and offloading decision results back to the MCC server. The MCC server updates the global optimization results and auxiliary variables using the closed-form solutions (32), (33), (41), and (58), and then dispatches these updates back to the respective MEC servers. Finally, when the iterative optimization is completed, the optimized beamforming vectors and offloading decision results are sent to the corresponding terminals.}}
\section{Simulation Results}

In this section, we present the simulation results to demonstrate the effectiveness of the proposed distributed CET-ISCC (DCET-ISCC) scheme. 

 \textcolor{black}{\subsection{System Setup and Benchmarking Schemes for Comparisons}}
The number of BSs is set as 3, which are located at (0m,0m), (400m,0m), and (200m,200$\sqrt{3}$), respectively; the 9 terminals are randomly distributed within a 1km$\times$1km two-dimensional area. Each terminal's sensing target is randomly located with the direction and distance generated by following a uniform distribution over [$0^\circ$, $180^\circ$] and [30m, 70m], respectively, and their RCS are assumed to be uniformly distributed between [0.8, 1]. Most of the simulation parameters are listed in Table \ref{table_example2}. According to \cite{chen2018} and \cite{wuiot}, the relationship between the CPU frequencies for MCC, MEC, and terminals  satisfies $f_{k}^C>f_{k}^E>f_{k}^L$. Moreover, we assume that all communication channels follow the Rayleigh distribution. \textcolor{black}{To obtain statistically accurate results, we do Monte Carlo simulation with the results being averaged over 500 trials. For comparisons,} we also provide the performance of the following four different schemes under the same parameter configuration.

\begin{table}[t]
	\renewcommand{\arraystretch}{1.3}
	\caption{SIMULATION PARAMETERS.}
	\label{table_example2}
	\centering
	\begin{tabular}{l l}
		\hline
		
		\bfseries Parameters &  \multicolumn{1}{c}{\bfseries Value}\\ 
		\hline
		
		Number of BSs $L$  & 3\\
		Number of terminals $K$  & 9\\
		Number of BSs' antennas $M$ & 16\\
		Number of terminals' antennas $N$  & 8\\
		Terminals' signal bandwidth $B$  & 10 MHz  \\
		The task computation intensity $\beta$ & 400 cycles/bit \\
		The numbers of bits of sensing task $Z_k$ & 100KB\\
		The path loss $\rho_0$ at the reference distance $d_0 = 1$ m   & -60 dB   \\
		\tabincell{l}{Terminal $k$'s CPU frequency} $f_{k}^L$  & 0.5 Gcycles/s\\
		\tabincell{l}{The CPU frequency allocated to  terminal $k$ by the \\ MEC $f_{k}^E$}& 3 Gcycles/s  \\
		\tabincell{l}{The CPU frequency allocated to terminal $k$ by the\\ MCC $f_{k}^C$}  & 10 Gcycles/s\\
		\tabincell{l}{The terminal's maximum power $P_{th}$} & 30 dBm\\
		The Gaussian noise power at BS $l$ $\sigma_b^2$ & -174 dBm/Hz\\
		The power coefficient related to the chip architecture $\eta$ & $10^{-28}$ \\
		\tabincell{l}{The transmission rate between MEC server and MCC\\ server $r^f$}  & 10 Mbit/s\\
		The maximum computation capacity of MEC \\ server $l$ ${G}_l$ & 9 Gcycles/s\\
		The sensing SINR threshold of terminals $\Gamma_{r}$ & 2 dB \\
		\hline
		
	\end{tabular}
\end{table}

\begin{itemize}
	
	\item {\bfseries{Edge-Terminal ISCC scheme (ET-ISCC) \cite{ISCC2}}}: For the Edge-Terminal ISCC scheme, terminals choose to offload the sensing tasks to the MEC server or process them locally. 
	\item {\bfseries{Local Execution scheme}}: For the local execution scheme, the sensing data collected by each terminal is processed locally.
	\item {\bfseries{Centralized CET-ISCC (CCET-ISCC) \cite{liu2024joint}}}:  In this scheme, all the channel state information is collected by a central server, which performs global offloading strategy and beamforming design, and then disseminates the results to BSs and terminals.
	\item {\bfseries{Distributed CET-ISCC with maximal ratio transmission (DCET-MRT)}}: In this scheme, terminals use the maximal ratio sensing method, i.e., using the ISAC beamforming vector $\mathbf{w}_k=\sqrt{P_{th}}\frac{\mathbf{h}_d}{||\mathbf{h}_{d}||}$, where $\mathbf{h}_{d}$ represents the $d$-th column of $\mathbf{H}_{lk}$ corresponding to the largest eigenvalue. 
	\item {\bfseries{Distributed CET-ISCC with maximal ratio sensing (DCET-MRS)}}: In this scheme, terminals adopt the maximal ratio sensing beamforming scheme, i.e., aligning the beam with the sensing target to achieve a higher echo SINR. The ISAC beamforming vector is represented as $\mathbf{w}_k=\sqrt{P_{th}}\frac{\mathbf{a}({\theta}_k)}{||\mathbf{a}({\theta}_k)||}$ . 
\end{itemize}

\begin{figure}
	\setlength{\abovecaptionskip}{-0.1 cm}
	\setlength{\belowcaptionskip}{-1cm}
	\centering
	\begin{subfigure}[]
		{\centering
			\includegraphics[width=1.8in]{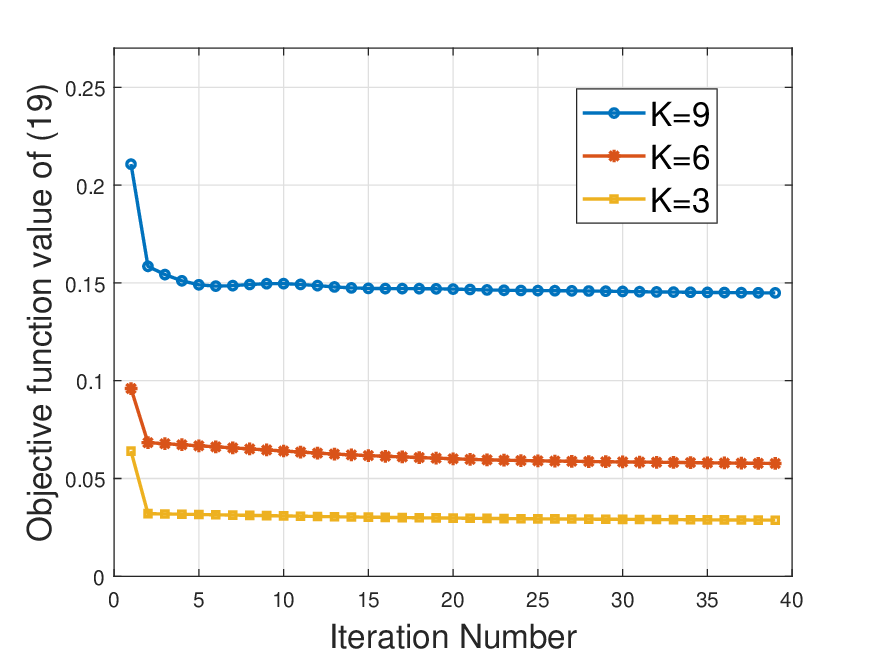}}
	\end{subfigure}
	\hspace{-7 mm}
	\begin{subfigure}[]
		{\centering
			\includegraphics[width=1.8in]{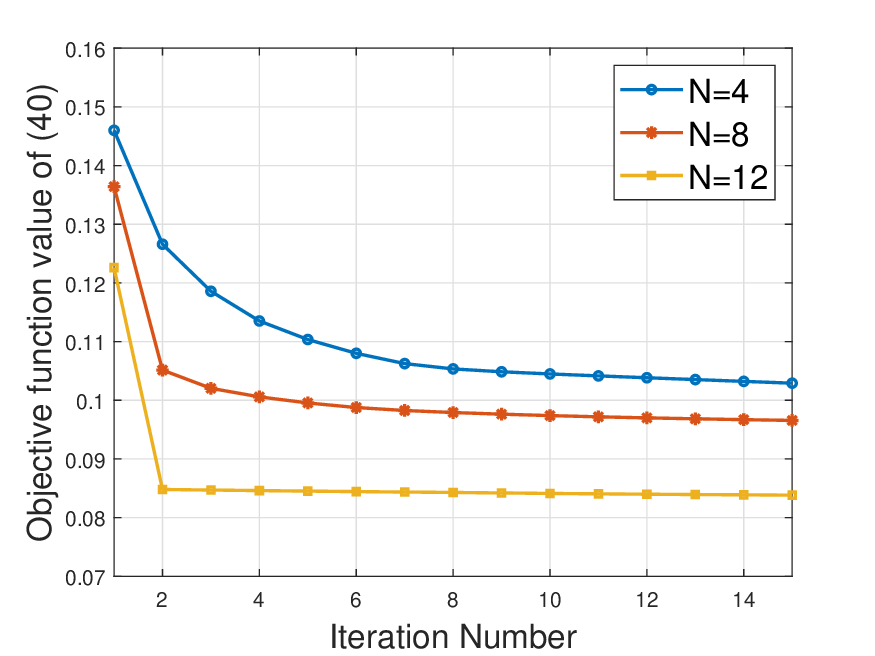}}
	\end{subfigure}
	\caption{(a) The convergence performance of Algorithm 1. (b) The convergence performance of Algorithm 2. }
	\label{fig:f1}
\end{figure}

 \textcolor{black}{\subsection{Convergence, computational complexity, and the average energy consumption performance}}

 \textcolor{black}{In this subsection, we analyze the convergence, computational complexity, and energy consumption performance of the proposed DCET-ISCC scheme.} We firstly investigate the convergence behavior of Algorithm 1 and Algorithm 2 in Fig.~\ref{fig:f1}(a) and Fig.~\ref{fig:f1}(a), respectively. As can be seen, both algorithms are able to converge in only a few iterations. Moreover, as the number of antennas increases, the degrees of freedom for ISAC beamforming optimization are increased, thereby reducing the objective value of problem (\ref{prob30}).

\begin{figure}[!t]
	\centering
	\includegraphics[width=3.2in]{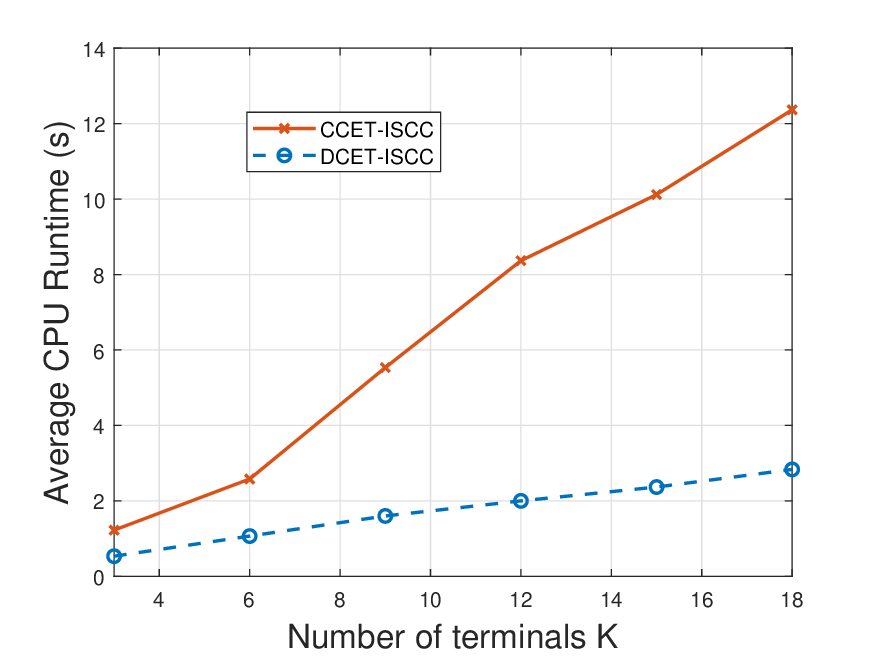}
	\caption{Average CPU runtime versus the number of terminals.}
	\label{fig:f8}
\end{figure}

To show the advancement of the DCET-ISCC scheme in lowering the computational complexity, in Table \ref{Tbl:Suspiciousness}, we compare the complexity performance of the distributed algorithm with that of the centralized algorithm. It is evident that the DCET-ISCC scheme has the lower computational complexity than the CCET-ISCC scheme. In Fig.~{\ref{fig:f8}}, we present the average CPU runtime of the two algorithms using MATLAB R2022 on  Intel Core i9-13900K. For the DCET-ISCC system, we use the Parallel Computing Toolbox in MATLAB for processing. It can be observed that the CPU runtime of the distributed algorithm is less than that of the centralized algorithm, and the reduction significantly increases with the growth in the number of terminals. This is because the offloading decision optimization and beamforming optimization are distributed across different BSs for parallel processing in the DCET-ISCC scheme, avoiding the concentration of computational load.

\renewcommand{\arraystretch}{2}

\begin{table}[!t]
	\centering
	\caption{Computational complexity comparison of different schemes}
	\label{Tbl:Suspiciousness}
	\renewcommand{\arraystretch}{1.5}
	\begin{tabular}{|c|c|}
		\hline
		Schemes &Total computational complexity
		\\ \hline
		CCET-ISCC\cite{liu2024joint}     & $\mathcal{O}\left(K^3L^3+K^{3}N^{3}+KM^3+KM^2N\right)$ \\  \hline 
		DCET-ISCC    & $\mathcal{O}\left(KL+K^3+K^{2}N^3+M^3+M^2N \right)$\\ 
		\hline
		
	\end{tabular}
\end{table}

\begin{figure}[!t]
	\centering
	\includegraphics[width=3.2in]{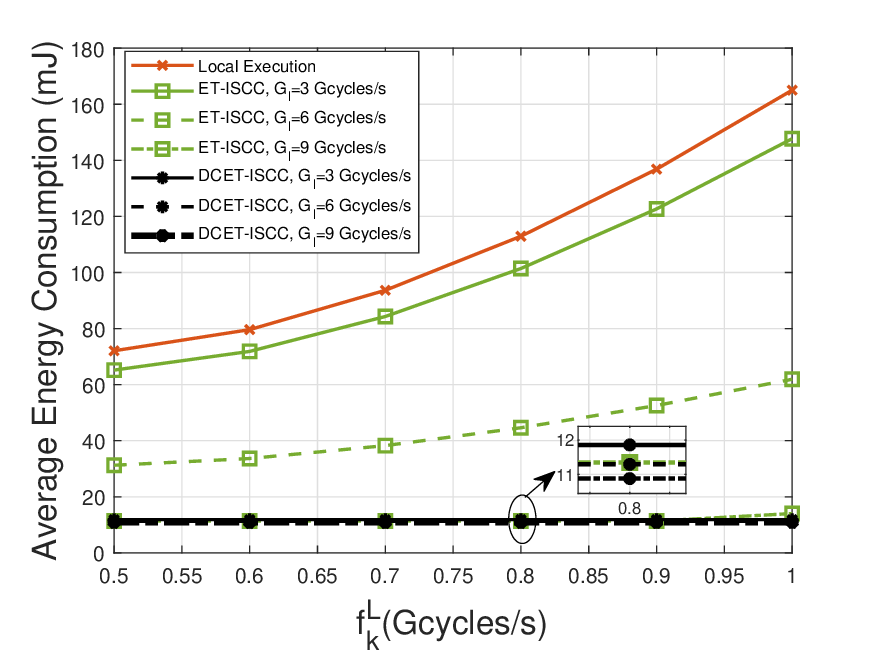}
	\caption{Average energy consumption versus different terminals' CPU frequency $f_k^L$.}
	\label{fig:f9}
\end{figure}

\textcolor{black}{In Fig.~\ref{fig:f9}, we demonstrate the terminals' average energy consumption $E_k=P_k^{total}T_k^{total}$ with different MEC servers' computation capacities $G_l$ and terminal CPU frequencies $f_k^L$. It can be observed that compared to the local execution and ET-ISCC schemes, the proposed DCET-ISCC scheme achieves lower energy consumption. The energy savings increase increase as the MEC servers' computation capacities $G_l$ decrease. This is because DCET-ISCC more effectively utilizes the cloud computing resources in the network, thereby significantly reducing the computational burden on terminals even when the computational capability of MEC servers is limited. Moreover, the energy consumption of the local execution and ET-ISCC schemes increases with the increase of $f_k^L$, while the DCET-ISCC scheme is almost unaffected. This stability is also due to the introduction of MCC servers, to which terminals can offload sensing tasks reducing their energy consumption.}

 \textcolor{black}{\subsection{The average sensing task execution latency  performance}}

\begin{figure}[!t]
	\centering
	\includegraphics[width=3.0in]{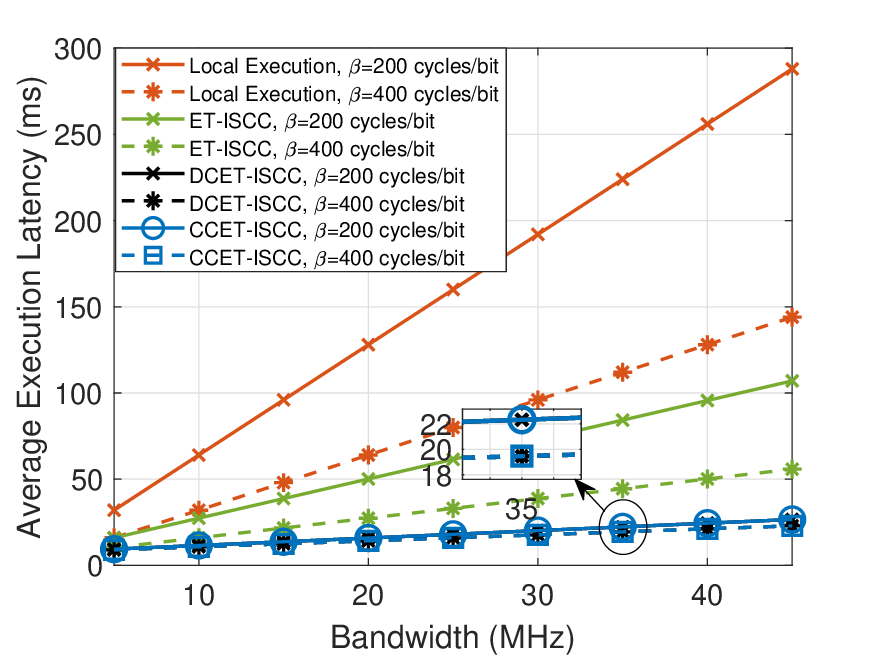}
	\caption{Average execution latency versus different signal bandwith $B$.}
	\label{fig:f3}
\end{figure}

\textcolor{black}{In this subsection, we further investigate the average sensing task execution latency  performance of the proposed DCET-ISCC scheme.} In Fig.~\ref{fig:f3}, we show the average sensing task execution latency of terminals versus the signal bandwidth under different offloading schemes. It can be seen that as the signal bandwidth increases, the average execution latency also increases. \textcolor{black}{This is because, according to the Nyquist sampling theorem, larger bandwidth results in a greater amount of echo signals to be processed. Although the increase of bandwidth also improves the offloading rate, keeping the offloading latency $t_k^{UP}$ almost the same, the linear increase in data volume leads to a linear increase in computing latency $t_k^{E}$, which is significantly higher than the offloading latency. Consequently, the average execution latency changes almost linearly with the increase of bandwidth.} Compared to the local execution scheme, the average execution latency in ET-ISCC and DCET-ISCC schemes grow more slowly with increasing bandwidth. This is because these two schemes can opt to offload computation-intensive tasks to MEC servers with higher computational capabilities. Additionally, the increase in bandwidth results in higher communication rate for offloading, thereby making the task execution latency less sensitive to bandwidth changes. Compared to the ET-ISCC scheme, the proposed DCET-ISCC scheme offers more diverse computing offloading options. When the computing capacities of the MEC servers are insufficient or the computing task is difficult to handle, they can further offload data to the MCC server for processing, resulting in better execution latency performance than ET-ISCC. Moreover, it shall be noted that the proposed distributed algorithm (DCET-ISCC) achieves the same performance as the centralized algorithm (CCET-ISCC), while at significantly reduced complexity. 

In Fig.~\ref{fig:f3}, we also provide the average execution latency performances under different task computation intensities $\beta$. It is observed that for the local execution and ET-ISCC schemes, the increase of $\beta$ significantly increases the average execution latency, while the proposed DCET-ISCC scheme is less affected. This is because, by leveraging the powerful computational capabilities of MCC server, DCET-ISCC can process high task computation intensity sensing tasks more efficiently. 

\begin{figure}[!t]
	\centering
	\includegraphics[width=3.0in]{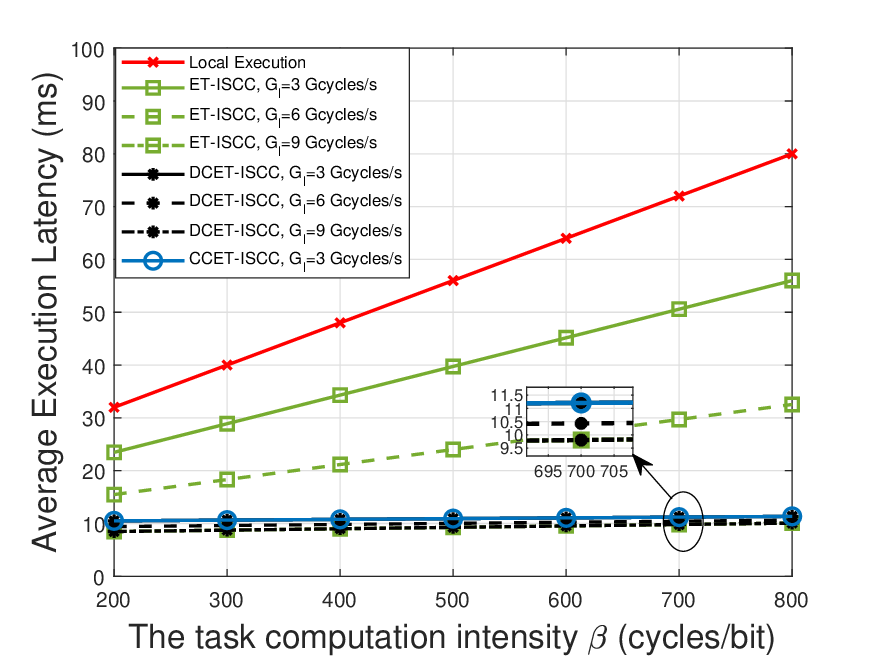}
	\caption{Average execution latency versus different task computation intensity $\beta$.}
	\label{fig:f4}
\end{figure}

To more intuitively demonstrate the superiority of the DCET-ISCC scheme in processing complex sensing tasks, we present the variation of the average execution latency with task computation intensities $\beta$ under different MEC servers' maximum computation capacities $G_l$ in Fig.~\ref{fig:f4}. It can be observed that under different $G_l$, the average execution latency of DCET-ISCC increases slowly with the increase of $\beta$, demonstrating its effectiveness in handling complex computational tasks. Additionally, it is noticeable that the performance of ET-ISCC is greatly influenced by $G_l$. For the case with  lower MEC servers' computation capacity, ET-ISCC has higher average execution latency, whereas DCET-ISCC, due to the presence of MCC server, is less affected by the MEC computation capacity constraint. When the computation capacity $G_l=9$G cycles/s, both the two schemes exhibit the same average latency performance. This is because the MEC server has sufficient computational resources to deal with the computation tasks of all terminals, eliminating the need of further offloading tasks to the MCC server.

\begin{figure}[!t]
	\centering
	\includegraphics[width=3.0in]{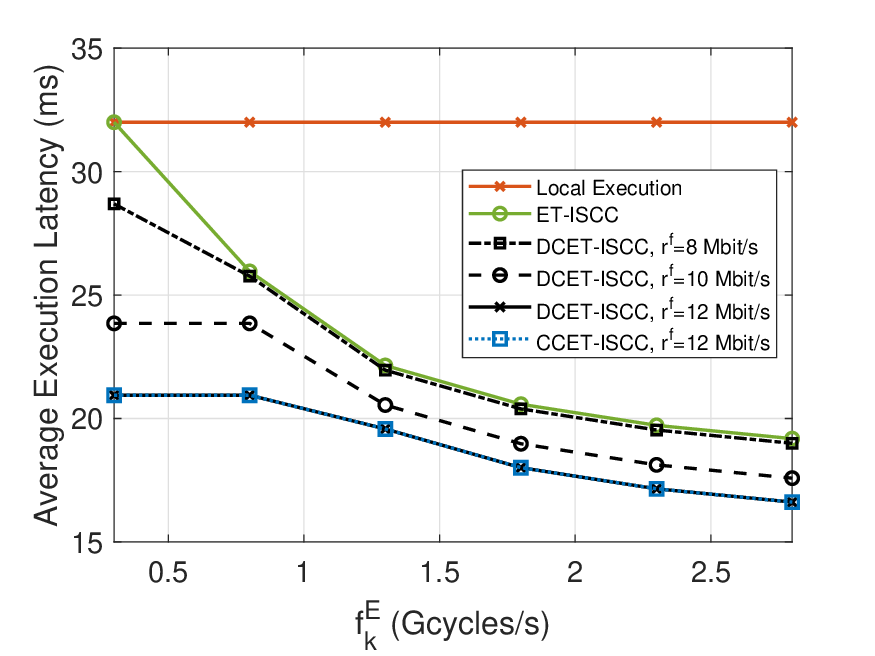}
	\caption{Average execution latency versus different CPU frequency $f_k^E$.}
	\label{fig:f5}
\end{figure}

To show the impacts of various factors on the execution latency, in Fig.~\ref{fig:f5},  we illustrate the average execution latency with different  CPU frequency $f_k^E$ allocated by MEC servers to terminals and MEC-to-MCC server transmission rate $r^f$. As can be seen, for all cases, the execution time of the DCET-ISCC time is shorter than that of the ET-ISCC scheme. Moreover, as the MEC-to-MCC server transmission rate decreases, the performances of both the ET-ISCC and DCET-ISCC schemes gradually overlap. This is because, for a small $r^f$, offloading tasks to the MCC server may induce a long transmission latency. The above results indicate that the offloading decision is influenced by various factors, and the proposed algorithm can select the optimal offloading strategy under different parameter settings.

\begin{figure}[!t]
	\centering
	\includegraphics[width=3.0in]{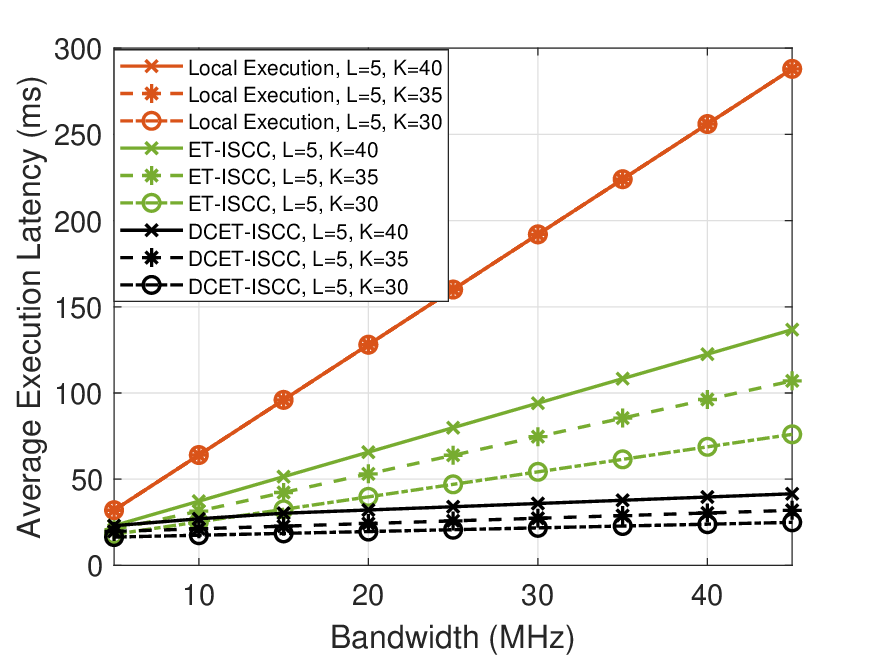}
	\caption{Average execution latency in larger-scale networks.}
	\label{fig:f10}
\end{figure}

\textcolor{black}{To further illustrate the effectiveness of the proposed DCET-ISCC scheme in large-scale networks, Fig.~\ref{fig:f10} shows the average execution latency with $L=5$ MEC servers and $K=30/35/40$ terminals. It can be observed that in larger-scale networks, the proposed three-tier architecture still outperforms the two-tier architecture. As the number of terminals increases, the multi-terminal interference also increases, resulting in a reduction in offloading rates and thereby increasing the average execution latency. Moreover, the gain of the three-tier architecture compared to the two-tier architecture becomes more pronounced as the number of terminals increases. This is because the three-tier system provides higher degrees of freedom for offloading, effectively mitigating the impact of insufficient computational resources in the two-tier architecture caused by an increase in terminal numbers.}

\begin{figure}[!t]
	\centering
	\includegraphics[width=3.0in]{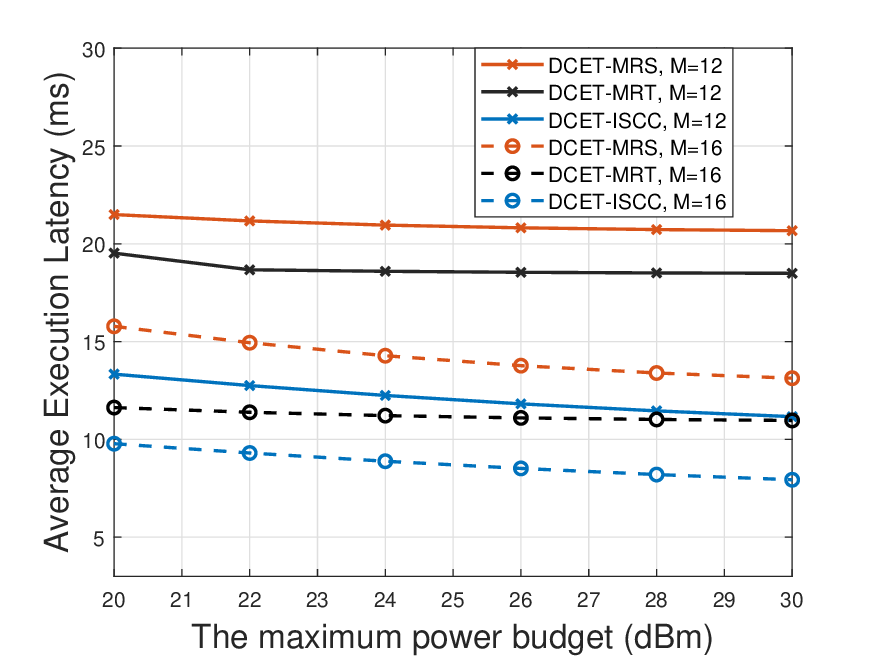}
	\caption{Average execution latency versus different maximum power budget of terminals $P_{th}$.}
	\label{fig:f6}
\end{figure}

Next, we investigate the impact of beamforming optimization on latency performance. In Fig.~\ref{fig:f6}, \textcolor{black}{we illustrate the average execution latency for different beamforming schemes under different maximum power budget of terminals $P_{th}$ and number of BS antennas $M$.} It can be observed that the proposed algorithm achieves a lower average execution latency compared to DCET-MRT and DCET-MRS. \textcolor{black}{This is because the optimized beamforming effectively eliminates the MTI among the terminals, which in turn increases the uplink transmission rate and significantly reduces the average execution latency.} Furthermore, we observe that the increase in $P_{th}$  has little impact on the performance of DCET-MRT and DCET-MRS, as the increase in $P_{th}$ leads to increased interference among terminals. In contrast, for the DCET-ISCC scheme, the degrees of freedom for optimization increase with the growth in $P_{th}$, resulting in a reduction of the average execution latency. Additionally, it is found that increasing the number of receiving antennas at BS is beneficial for all schemes due to the additional diversity gain.

\begin{figure}[!t]
	\centering
	\includegraphics[width=3.0in]{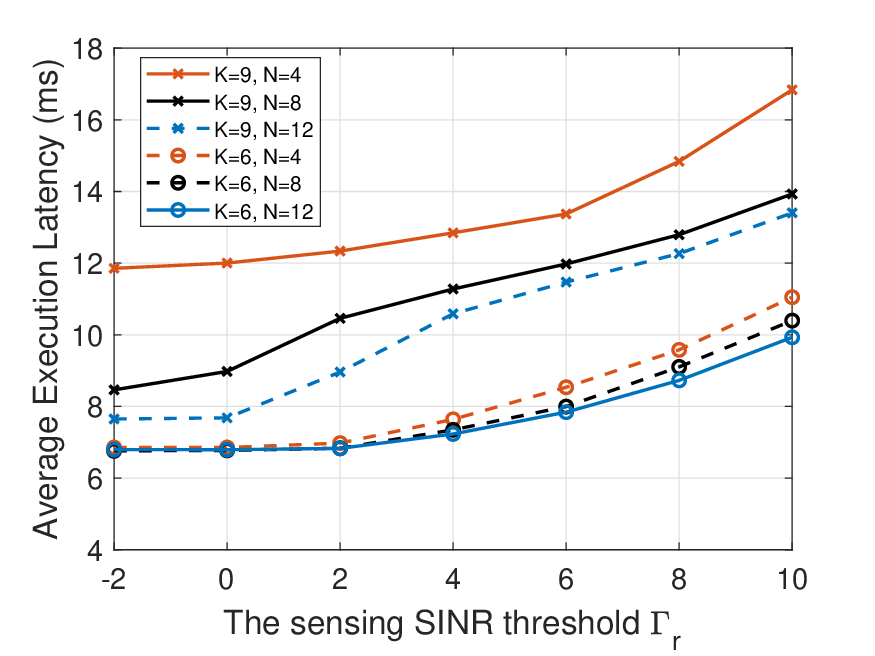}
	\caption{Average execution latency versus sensing SINR threshold $\Gamma_{r}$.}
	\label{fig:f7}
\end{figure}

\begin{figure}[!t]
	\centering
	\includegraphics[width=3.0in]{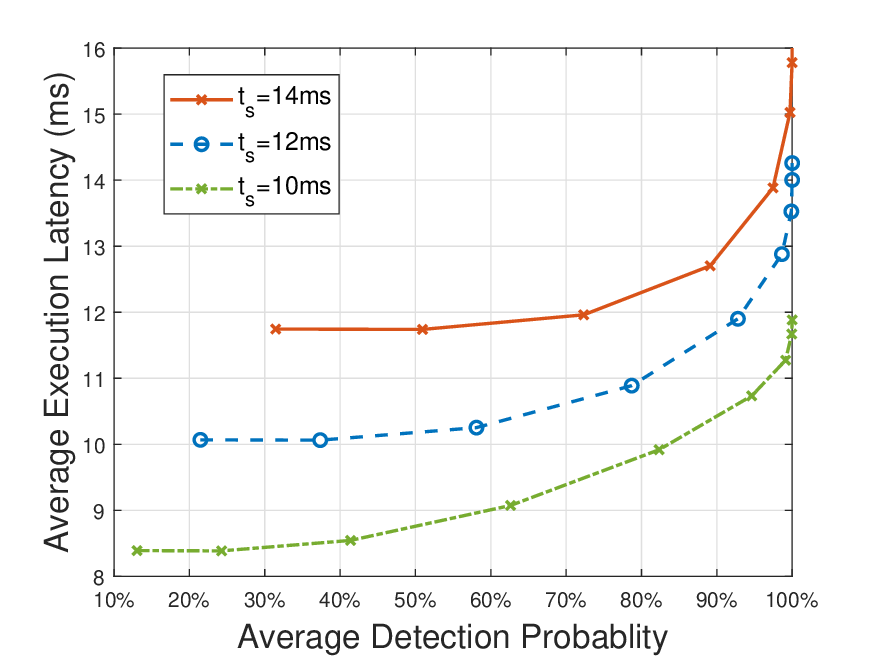}
	\caption{Average execution latency versus Averagey detection probablity.}
	\label{fig:f14}
\end{figure}

\begin{figure}[!t]
	\centering
	\includegraphics[width=3.0in]{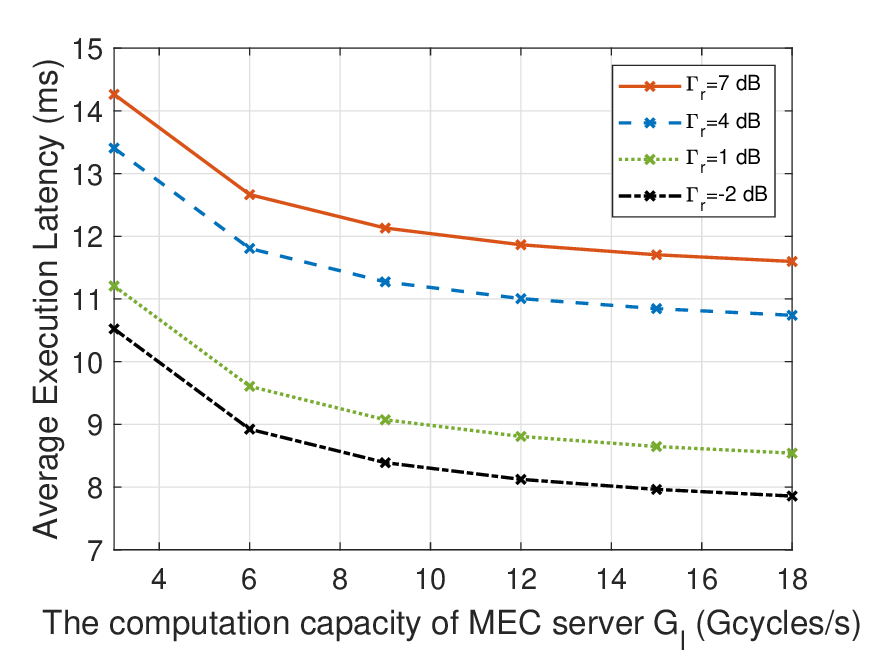}
	\caption{Average execution latency versus different MEC server's computation capacity $G_l$.}
	\label{fig:f12}
\end{figure}
Fig.~\ref{fig:f7} shows the impact of the sensing SINR threshold $\Gamma_{r}$ on the average execution latency of the DCET-ISCC scheme. As can be observed, as $\Gamma_{r}$ increases, the average execution latency also increases. This is because the higher sensing requirements compel the terminal to focus more power towards the sensing target, thereby resulting in a reduction in the uplink transmission rate. This phenomenon also reflects the trade-off between the sensing and communication functionalities in the ISCC  system. Additionally, Fig.~\ref{fig:f7} also illustrates the impact of the number of terminal antennas $N$ and the number of terminals $K$ on the system's latency performance. It can be seen that the increase of the number of terminal antennas $N$ results in an increased uplink transmission rate, which in turn reduces the average execution latency when the sensing SINR threshold remains constant. Meanwhile,  as the interference between terminals significantly decreases, the decrease in the number of terminals also reduces the  average execution latency. Moreover, with more terminals, the increase of the number of antennas brings greater performance gains. This is because the increased interference caused by a larger number of terminals can be significantly mitigated by increasing the number of antennas, thereby enhancing the system latency performance.

\textcolor{black}{To further reveal the trade-off between sensing performance and data offloading,  we simulate the variation  in average execution latency with respect to the average detection probability under different sensing echo accumulation times $t_s$ using the Neyman-Pearson criterion and the Generalized Likelihood Ratio Test \cite{1998dec} \cite{rihan2023}. As shown in Fig.~\ref{fig:f14}, we observe that as the average detection probability increases, the average execution latency also rises. Additionally, an increase in the sensing echo accumulation time $t_s$ effectively enhances the average detection probability; however, it also leads to an increase in the volume of sensing data, which in turn raises the average execution latency for sensing tasks. The above results clearly demonstrate the trade-off between computational latency and sensing performance.}

\textcolor{black}{In Fig.~\ref{fig:f12}, we present the average execution latency under different radar sensing thresholds $\Gamma_r$ and MEC computational capacities $G_l$. With the increase of MEC computational resources, the average execution latency for sensing tasks \textcolor{black}{accordingly} decreases. Additionally, we can observe that the decreasing speed of execution latency reduces with the computation capability of MEC servers. This is because the execution latency depends on both computation capability of MEC servers and offloading rate. When computational resources are ample, the average execution latency is primarily influenced by the offloading latency. Furthermore, under a fixed average execution latency, achieving a higher sensing SINR requires more computational resources from the MEC server, \textcolor{black}{reflecting} the trade-off between MEC computational resources and sensing performance within a three-tier ISCC architecture.}
\section{Conclusion}
In this paper, we investigated joint optimization of computation offloading and beamforming design in a three-tier ISCC system, where terminals can optionally offload sensing tasks to MEC servers or MCC server while performing radar sensing. To reduce the average execution latency of sensing tasks and computational complexity in joint design, we proposed a distributed optimization framework based on the ADMM and WMMSE approach to optimize the computation offloading and ISAC beamforming subproblems, respectively, in an alternating manner. Simulation results demonstrate the convergence and effectiveness of the proposed distributed algorithm. Compared to existing two-tier ISCC  and local execution schemes, our proposed three-tier architecture offers higher DoFs in offloading, thereby effectively reducing average execution latency. Moreover, the proposed distributed algorithm significantly reduces algorithm runtime compared to centralized approach. The impacts of different parameters, including antenna number and MEC server computation capacity on latency \textcolor{black}{and energy consumption} performance have been investigated, revealing the trade-off between task offloading efficiency and sensing accuracy.

\textcolor{black}{Although this paper demonstrates  the superiority of the three-tier ISCC architecture, there are still many challenges regarding deploying and implementing this architecture in the real world. For the future research, the following directions can be considered:}
\begin{itemize}
	\item \textcolor{black}{\textbf{Protocol Design:}: In practical network deployments, the proposed scheme requires robust information exchange among MEC servers, MCC servers, and ISAC terminals. Developing and refining protocols to ensure efficient and reliable information exchange is crucial for the implementation in the real-world.}
	
	\item \textcolor{black}{\textbf{Intelligent Processing of Sensing Data}: Introducing Artificial Intelligence has the potential to revolutionize the processing of large volumes of raw sensing data collected by MEC and MCC servers in the three-tier ISCC framework. Developing efficient AI algorithms to effectively handle and interpret this data remains a significant challenge and a promising direction for further research.}
	
	\item \textcolor{black}{\textbf{Security Issues}: Sensing data offloading over wireless links to MEC servers introduces potential security risks. Ensuring secure data transmission within the ISCC three-tier architecture is imperative. To achieve this, it is required to develop robust physical layer security measures and encryption techniques to protect data from being overheard.}
\end{itemize}


\appendix
\section*{Appendix A: Proof of Lemma 1}
With fixed $\mathbf{u}_k$ and $\mathbf{w}_k$, the optimal $V_k$ can be obtained through the first-order optimality condition, i.e.,
\begin{align} 
\label{gradient}
\frac{\partial \widehat{R}_{lk}(V_k)}{\partial V_k}=(V^{-1}_k)^T-E^T_k=0,
\end{align}
and the optimal $V_k$ is given by
\begin{equation}
V^{opt}_k=E_k^{-1}=(1-\mathbf{w}^H_{k}\mathbf{H}_{lk}({\sigma_b}^2\mathbf{I}_M+\mathbf{F}_k)^{-1}\mathbf{H}^H_{lk}\mathbf{w}_{k})^{-1}.
\end{equation}

By substituting the optimal $V_k^{opt}$ in (\ref{MMSE2}) into (\ref{r1u}), $\widehat{R}_{lk}$ in (\ref{r1u}) can be reformulated as
\begin{equation}\small
\begin{aligned}
\label{r3}
&\widehat{R}_{lk}=B(\log(V_k)-V_kE_k+1)=B(\log(E^{-1}))\\
&=B(\log((1-\mathbf{w}^H_{k}\mathbf{H}_{lk}({\sigma_b}^2\mathbf{I}_M+\mathbf{F}_k)^{-1}\mathbf{H}^H_{lk}\mathbf{w}_{k})^{-1}))\\
&=B(\log(1+\mathbf{w}^H_{k}\mathbf{H}_{lk}({\sigma_b}^2\mathbf{I}_M+\mathbf{F}_k-\mathbf{H}^H_{lk}\mathbf{w}_{k}\mathbf{w}^H_{k}\mathbf{H}_{lk})^{-1}\mathbf{H}^H_{lk}\mathbf{w}_{k}))\\
&=B(\log(1+\mathbf{w}^H_{k}\mathbf{H}_{lk}({\sum_{i=1,i\neq k}^K\mathbf{H}_{li}^{H}\mathbf{w}_{i}\mathbf{w}^H_{i}\mathbf{H}_{li}}+{\sigma_b}^2\mathbf{I}_M)^{-1}\mathbf{H}^H_{lk}\mathbf{w}_{k}))\\
&=B\log(1+\mathbf{w}_{k}^H\mathbf{H}_{lk}\mathbf{D}^{-1}_{lk}\mathbf{H}_{lk}^{H}\mathbf{w}_{k}),
\end{aligned}
\end{equation}
\textcolor{black}{where the third equality is obtained via the Woodbury matrix identity.}

On the other hand, based on the property that $\det(\mathbf{I}+\mathbf{AB})=\det(\mathbf{I}+\mathbf{BA})$, (\ref{r1}) can be rewritten as
\begin{equation}
\begin{aligned}
\label{r4}
R_{lk} =& B\log\det(\mathbf{I}_M+\mathbf{H}_{lk}^{H}\mathbf{w}_{k}\mathbf{w}_{k}^H\mathbf{H}_{lk}\mathbf{D}^{-1}_{lk})\\
=&B\log(1+\mathbf{w}_{k}^H\mathbf{H}_{lk}\mathbf{D}^{-1}_{lk}\mathbf{H}_{lk}^{H}\mathbf{w}_{k}),
\end{aligned}
\end{equation}

Combining (\ref{r3}) with (\ref{r4}), we complete the proof of lemma 1.

\appendix
\section*{Appendix B: Proof of Lemma 2}
For ease of illustration, we denote $\bm{\Upsilon}$ in problem (\ref{prob23}) as $\widehat{\bm{\Upsilon}}$. Since $\{\mathbf{w}_{k}\}^K_{k=1}$ is obtained by solving equation (\ref{prob23}), they satisfy $\text{tr}({{\mathbf{H}}_{jk}^{IH}}\mathbf{w}_{k}\mathbf{w}^H_{k}{\mathbf{H}}^{I}_{jk})\leq\widehat{\Upsilon}_{jk}, \forall k, \forall j\neq k$. When $\Upsilon^*_{jk}=\text{tr}({{\mathbf{H}}_{jk}^{IH}}\mathbf{w}_{k}\mathbf{w}^H_{k}{\mathbf{H}}^{I}_{jk}),\forall k, \forall j\neq k$, we have
\begin{equation}
\begin{aligned}
\label{r5}
\frac{\Gamma_{k}}{\xi_k^2}{\sum_{j=1,j\neq k}^K\Upsilon^*_{kj}}+\sigma_k^2&=\frac{\Gamma_{k}}{\xi_k^2}{\sum_{j=1,j\neq k}^K\text{tr}({{\mathbf{H}}_{kj}^{IH}}\mathbf{w}_{j}\mathbf{w}^H_{j}{\mathbf{H}}^{I}_{kj})}+\sigma_j^2\\
&\leq\frac{\Gamma_{k}}{\xi_k^2}{\sum_{j=1,j\neq k}^K\widehat{\Upsilon}_{kj}}+\sigma_k^2.
\end{aligned}
\end{equation}

Since $\{\mathbf{w}_{k}\}^K_{k=1}$ and $\widehat{\bm{\Upsilon}}$ satisfies (\ref{prob3}), we have
\begin{equation}
\begin{aligned}
\label{prob3u} 
&\frac{\Gamma_{k}}{\xi_k^2}{\sum_{j=1,j\neq k}^K\widehat{\Upsilon}_{kj}}+\sigma_k^2  \\ \leq&2\text{tr}(\widetilde{\mathbf{w}}^H_{k}\mathbf{A}({\theta}_k)^H\mathbf{A}({\theta}_k)\mathbf{w}_{k})-\text{tr}(\widetilde{\mathbf{w}}^H_{k}\mathbf{A}({\theta}_k)^H\mathbf{A}({\theta}_k)\widetilde{\mathbf{w}_{k}}).
\end{aligned}
\end{equation}

Combining (\ref{r5}) with (\ref{prob3u}), $\{\mathbf{w}_{k}\}^K_{k=1}$ and $\bm{\Upsilon}^*$ satisfy (\ref{prob3}). Additionally, $\Upsilon^*_{jk}=\text{tr}({{\mathbf{H}}_{jk}^{IH}}\mathbf{w}_{k}\mathbf{w}^H_{k}{\mathbf{H}}^{I}_{jk}),\forall k, \forall j\neq k$ satisfy constraints (\ref{prob24}a). To sum up, (\ref{gamaup}) is a feasible solution to problem (\ref{prob24}), and Lemma 2 is proven.

\bibliographystyle{IEEEtran}
\bibliography{bibfilelp2}

\end{document}